\documentclass{iopart}
\usepackage{graphicx}
\usepackage[utf8]{inputenc}
\usepackage{wrapfig}
\usepackage{blindtext} 
\usepackage{todonotes}
\usepackage{iopams}  
\usepackage{comment}

\usepackage{mathtools}
 \usepackage{booktabs} 
\usepackage{float}
\usepackage{csquotes}
\usepackage{subfigure}
\usepackage{booktabs}
\usepackage{adjustbox}
\usepackage{amsmath}
\usepackage{tipa}
\usepackage{lmodern}
\usepackage{fancyhdr}
\usepackage[textwidth=6in]{geometry}
\usepackage{multicol}
\usepackage{multirow}
\usepackage[inline]{enumitem}
\usepackage{lipsum}
\usepackage{adjustbox}
\usepackage{natbib}
\usepackage[english]{babel}
\usepackage[colorlinks=true, citecolor=blue, linkcolor=blue, urlcolor=blue]{hyperref}

\usepackage[section]{placeins}
\usepackage[pagewise]{lineno}

\begin{document}
\title{ Mapping of  discrete range modulated proton radiograph to water-equivalent path length using machine learning}

\author{Atiq~Ur~Rahman$^1$,  Chun-Chieh Wang $^1$$^2$$^3$, Shu-Wei Wu $^1$$^4$,  Tsi-Chian~Chao$^1$$^2$$^5$$^6$$^7$ \footnote{Equal contribution}, I-Chun~Cho$^3$$^7$ \footnotemark[1]}
\address{Institute}
\address{$^1$  Department of Medical Imaging and Radiological Sciences, College of Medicine, 
No.259, Wenhua 1st Rd., Guishan Dist., Taoyuan City, 333323, Taiwan}

\address{$^2$ Department of Radiation Oncology, Linkou Chang Gung Memorial Hospital, No.15, 
Wenhua 1st Rd., Guishan Dist., Taoyuan City, 333011, Taiwan }

\address{$^3$ Radiation Research Core Laboratory, Linkou Chang Gung Memorial Hospital, No.15, 
Wenhua 1st Rd., Guishan Dist., Taoyuan City, 333011, Taiwan }

\address{$^4$ Proton and Radiation Therapy Center, Department of Radiation Oncology, Linkou Chang Gung Memorial Hospital, Taoyuan City, 333, Taiwan}

\address{$^5$ Center for Reliability Sciences and Technologies, Chang Gung University, No.259, 
Wenhua 1st Rd., Guishan Dist., Taoyuan City, 333323, Taiwan}

\address{$^6$ Department of Radiation Oncology, New Taipei Municipal Tucheng Hospital, No.6, 
Sec.2, Jincheng Rd., Tucheng Dist., New Taipei City 236043, Taiwan }

\address{$^7$ Research Center for Radiation Medicine, Chang Gung University, No.259, Wenhua 1st 
Rd., Guishan Dist., Taoyuan City, 333323, Taiwan }

\ead{\href{mailto:iccho@mail.cgu.edu.tw}{iccho@mail.cgu.edu.tw}}
\ead{\href{mailto:chaot@mail.cgu.edu.tw}{chaot@mail.cgu.edu.tw }}

\vspace{10pt}
\begin{indented}
\item[] \today
\end{indented}
\noindent\rule{\textwidth}{0.4pt}
\begin{abstract}
Objective. Proton beams enable highly localized dose delivery in radiotherapy due to their distinct energy deposition profiles. Accurate proton range estimation is vital for proton therapy. However, current treatment planning relies on X-ray CT, which introduces uncertainties in stopping power conversion and range calculation. Proton CT offers a more direct, physics-based method by measuring water-equivalent thickness, but its clinical use has been hampered by spatial resolution loss due to multiple Coulomb scattering. This study aims to overcome these limitations by developing a direct, data-driven method for reconstructing water-equivalent path length maps from energy-resolved proton radiographs, thereby bypassing intermediate reconstruction steps and inherent sources of error in conventional workflows. Approach. This work presents a machine learning pipeline for  WEPL reconstruction from high-dimensional, proton radiograph. Data was generated using the TOPAS Monte Carlo toolkit, modeling a clinical proton therapy nozzle and a patient CT dataset, with proton energies spanning 70--230~MeV and 72 projection angles. Principal component analysis was used to reduce the input dimensionality of the radiographs. A conditional  generative adversarial network with gradient penalty was trained for WEPL prediction. Combined adversarial loss, mean squared error, structural similarity, and perceptual terms  used to improve  quality and model stability. Main results. The proposed model achieved  mean relative WEPL deviation of 2.5\%, a structural similarity of 0.97, and a proton radiography gamma index passing rate of 97.1\% ( 2\% $\Delta$WEPL, 3~mm distance-to-agreement criteria) on simulated head phantom, demonstrating high accuracy and fidelity in both spatial and structural features. Significance. This study demonstrates the feasibility of  WEPL mapping from proton radiographs using deep learning, bypassing intermediate reconstruction steps. The method addresses limitations of analytic techniques and paves the way for improved treatment planning. 
Future directions include optimizing PCA component selection, incorporating  detector response, exploring low-dose scenarios, and extending multi-angle data to proton CT reconstruction.
\end{abstract}

\noindent\rule{\textwidth}{0.4pt}

\section{Introduction}\label{section:intro}
Proton therapy is an advanced radiation modality that delivers highly localized tumor doses while minimizing damage to surrounding healthy tissues and sparing organs at risk (OAR). However, its clinical effectiveness remains constrained by uncertainties in proton range calculation within the treatment planning system (TPS), arising from factors such as organ motion, patient setup, anatomical variations, and conversion errors from Hounsfield units in computed tomography (CT) images to tissue stopping power. To compensate for this uncertainty, safety margins for critical organs are typically set to 2--4\% of the total proton range plus 1--3~mm~\citep{Paganetti_2012}. Over the past few decades, multiple strategies have been explored to address range uncertainty in proton therapy, including dual-energy computed tomography, prompt-gamma range verification, in-beam positron emission tomography imaging, ionoacoustic tomography, and proton radiography~\citep{Hunemohr2014,Min2006,Knopf2013,Kellnberger2016}.

A major source of proton range uncertainty arises from patient imaging. Currently, proton therapy treatment plans rely on X-ray CT scans, where Hounsfield units provide electron density and elemental composition. The widely used stoichiometric method of Schneider \textit{et al.} converts Hounsfield units to relative proton stopping power through calibration curves~\citep{Schneider1996,Schneider2005}, but this conversion is inherently imperfect due to the lack of a direct correspondence between X-ray attenuation and proton stopping power~\citep{petzoldt2021ct}. To improve accuracy, methods such as dual-energy CT (DECT) have been explored, by employing multiple X-ray spectra to better characterize tissue and reduce stopping-power errors~\citep{charyyev2022synthetic}. Bäumer \textit{et al.}~\citep{Baumer_2021} also proposed ToF-PET-derived $\gamma$CT imaging to refine stopping-power estimation, but their approach suffers from reduced spatial resolution and increased noise compared to conventional CT, limiting its value for patient-specific calculations. Importantly, these techniques do not directly measure proton interactions, so their accuracy remains limited.

A more direct approach to reducing range uncertainty is proton imaging, including proton radiography (2D) and proton computed tomography (3D). This technique transmits high-energy, low-fluence protons through the patient and measures residual intensity downstream~\citep{koehler1968protonS}. Koehler \textit{et al.} pioneered proton radiography, demonstrating high sensitivity to density variations at low dose, though early approaches were limited by image blurring from multiple Coulomb scattering and were mainly suitable for thin, homogeneous samples. Testa \textit{et al.} used time-resolved dose measurements with a 2D diode-array detector to enable rapid, low-dose water-equivalent path length (WEPL) imaging and real-time tumor tracking, but resolution was constrained by detector limitations and lack of scattering correction~\citep{testa2013proton}. Proton imaging directly measures the energy loss or residual range, providing information on WEPL, with stopping-power ratio uncertainties of 1.6\% for soft tissue, 2.4\% for bone, and 5.0\% for lung~\citep{Yang2012}. Although proton imaging was developed in the 1960s, clinical adoption has been slow due to image quality issues and advances in diagnostic X-ray CT~\citep{Hanson1982}. Bentefour \textit{et al.} introduced energy-resolved proton radiography, using \( R_{80} \) (the depth at which dose falls to 80\% of its peak) from the proximal fall-off of the energy-resolved dose function (ERDF) to avoid image degradation due to multiple Coulomb scattering. By delivering sequential monoenergetic proton beams to a thin, pixelated detector, a stack of ERDFs is obtained. Extracting \( R_{80} \) from each pixel’s ERDF yields a spatially resolved 2D WEPL map, as shown in~\citep{Bentefour2016}. Similarly, proton cone-beam CT (PCB-CT) utilizes scattered proton beams and depth-dependent modulation for stopping-power estimation, with WEPL reconstruction using the Feldkamp–Davis–Kress (FDK) algorithm. While this improves accuracy, spatial resolution is still limited by multiple Coulomb scattering, emphasizing the need for advanced correction methods~\citep{Zygmanski_2000}.

\begin{figure}[!htbp]
    \centering
    \fbox{%
        \begin{minipage}{0.85\linewidth}
            \centering
            \includegraphics[trim={0cm 0cm 0cm 0cm},clip,scale=.55]{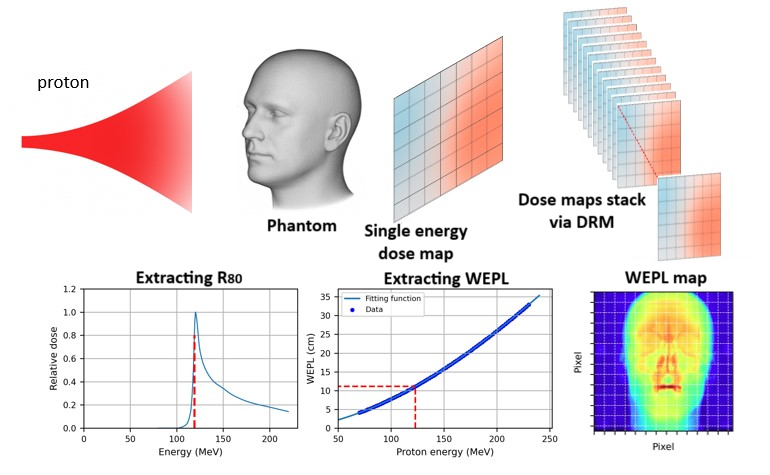}
            \caption{Workflow for discrete range modulation based WEPL mapping. 
                Top: Proton beam irradiation of a head phantom produces dose maps, DRM gives stacked of dose maps. 
                Bottom: For each pixel, the Bragg peak is analyzed to extract $R_{80}$, which is converted to WEPL (mm) using an energy-to-range calibration. The resulting values are assembled into a 2D WEPL map.
            }
            \label{fig:Abstract1}
        \end{minipage}
    }
\end{figure}

\begin{figure}[!htbp]
    \centering
    \fbox{%
        \begin{minipage}{0.85\linewidth}
            \centering
            \includegraphics[trim={0cm 0cm 0cm 0cm},clip,scale=.50]{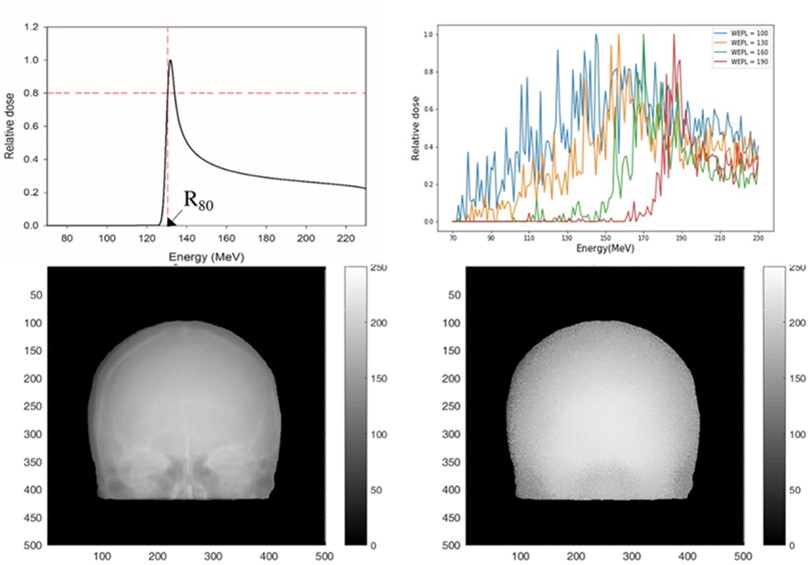}
            \caption{Impact of scattering and uncertainties on Bragg peak quality. 
            Top left: Ideal Bragg peak with sharp distal fall-off. 
            Top right: Distorted Bragg peaks due to noise and range straggling. 
            Bottom: Proton CT images reconstructed with clear (left) and blurred (right) Bragg peak resolution.}
            \label{fig:ScatteringImpact}
        \end{minipage}
    }
\end{figure}

A major study~\citep{Tsai2022DRM} investigated proton radiography using the discrete range modulation (DRM) method to improve WEPL accuracy in proton therapy as shown in Figure~\ref{fig:Abstract1}. Monte Carlo simulations with MCNPX2.7.0 compared DRM and Continuous Range Modulation (CRM) using parallel proton beams from 70 to 230~MeV in 2.5~MeV increments. DRM showed superior accuracy, reducing range mixing artifacts, achieving WEPL deviation within 2~mm or 1\%, and reaching a gamma passing rate of 0.999 in complex phantoms. Despite these advantages, image quality in energy-resolved proton radiography is still affected by the number of detected particles and structural complexity of the target. When proton beam counts are reduced to minimize radiation exposure, stochastic fluctuations in the ERDF arise. Multiple Coulomb scattering further distorts the ERDF in complex structures such as head phantoms. These factors make \( R_{80} \) determination more difficult, which decreases WEPL estimation accuracy, and limit image resolution, as shown in Figure~\ref{fig:ScatteringImpact}. Continued refinements are needed for clinical application.
In a related study, Cho \textit{et al.} demonstrated that DRM proton radiography can provide high spatial resolution and accurate WEPL prediction by correcting edge blurring from multiple Coulomb scattering. Their in-house deconvolution method, validated through simulations and clinical experiments, improved image sharpness and material boundary definition. However, most results were based on phantom data and require manual approaches. Further research is required to address patient motion and detector limitations~\citep{cho2025experimental}.

Traditional methods for estimating WEPL are based on analytical or iterative reconstruction from proton radiographs or pCT images~\citep{landry2018current}. These methods are often limited by low spatial resolution, noise sensitivity, and computational inefficiency, all of which can reduce the accuracy of WEPL predictions~\citep{parodi2020latest}. Previous DRM-based work by Chou \textit{et al.} demonstrated that WEPL can be estimated from energy-resolved proton radiography using discrete energy modulation and $R_{80}$-based methods. However, these approaches are constrained by range straggling and uncertainties from multiple Coulomb scattering, which compromise the precision of $R_{80}$ localization, as illustrated in Figure~\ref{fig:ScatteringImpact}. In anatomically complex regions, spatial uncertainty from scattering can further obscure the Bragg peak fall-off, resulting in degraded WEPL resolution.

Recent advances in machine learning (ML) and deep learning (DL) have enabled direct image reconstruction in medical imaging, bypassing conventional reconstruction chains~\citep{Andrew9161006,Rahman_2022}. Several DL-based approaches for proton CT reconstruction have been explored. In one such approach, Bayesian Convolutional Neural Network (BCNN) has been developed to correct noisy proton stopping power ratio (SPR) images from Monte Carlo simulations, yielding improved accuracy and uncertainty estimates for proton therapy planning~\citep{Nomura2021}. Similarly, Landry \textit{et al.} incorporated physics-informed priors into learning-based models, resulting in higher quality WEPL reconstructions for proton radiography~\citep{Landry2024}.

Despite advances in AI and computational imaging, reconstructing high-dimensional radiograph data with DRM remains challenging due to the large number of dose maps per projection, which limits the implementation of conventional deep learning (DL) approaches. High-dimensional images require significant computational and training resources. Large models can also suffer from underfitting, overfitting, and poor generalization, a consequence of the curse of dimensionality. Deep networks exposed to such data may struggle to extract meaningful features, often mistaking noise or irrelevant patterns for signal. As a result, conventional DL methods may only succeed as brute-force solutions that demand extensive computing power and large datasets.

\begin{figure}[!htbp]
\centering
\fbox{%
\begin{minipage}{0.85\linewidth}
            \centering
            \includegraphics[trim={0cm 0cm 0cm 0cm},clip,scale=.0850]{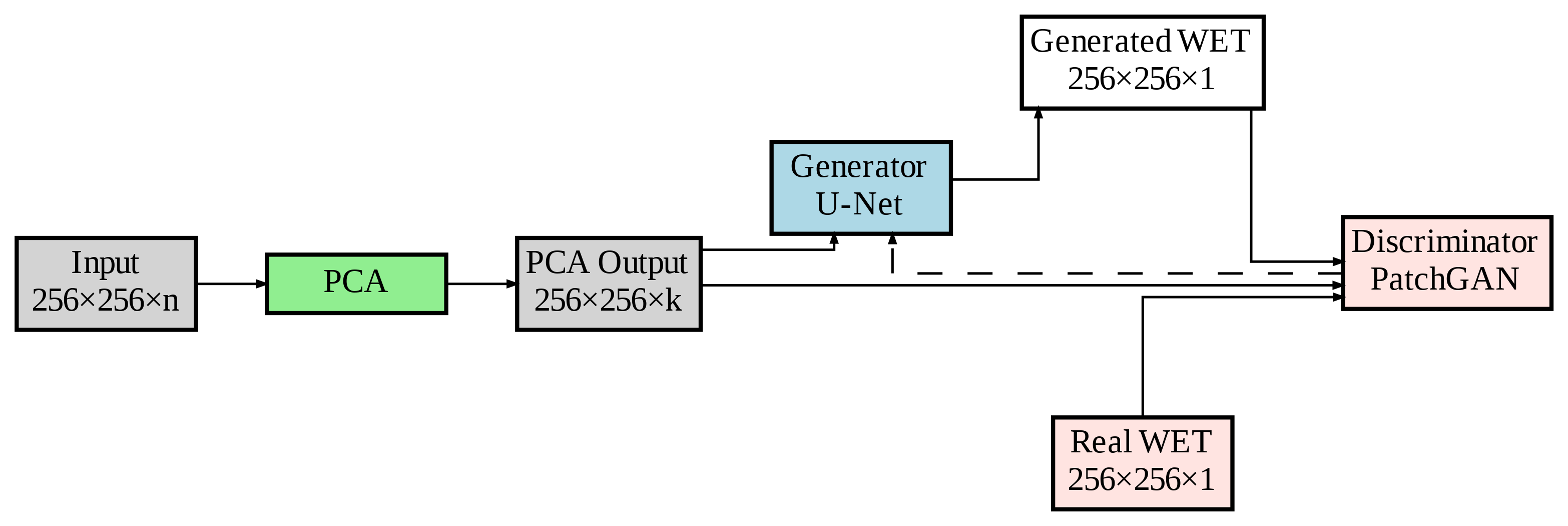}
            \caption{Block diagram of the PCA-conditioned cGAN architecture used for direct WEPL map synthesis from compressed proton radiograph data.}
            \label{fig:AbstractDia3}
        \end{minipage}
    }
\end{figure}

To address these challenges, we propose a direct AI-based approach for mapping WEPL maps from radiograph data, bypassing conventional reconstruction and mitigating the curse of dimensionality. Our method integrates principal component analysis (PCA) driven feature selection with a deep learning pipeline, as shown in Figure~\ref{fig:AbstractDia3}. Radiograph stacks are compressed using PCA, and the reduced representation serves as a conditional input for both the U-Net generator and PatchGAN discriminator. The generator produces WEPL maps, which are evaluated by the discriminator together with the real maps and PCA condition. PCA has demonstrated effectiveness in dimensionality reduction for CT, PET, MRI, and radiotherapy planning, supporting diagnostic integrity~\citep{ARGOTAPEREZ202213,realtimeMRrecopca,jimaging8020017,s23239418}. Although PCA is widely used in medical imaging, it has not been applied to reconstruct WEPL maps from proton radiography data. This study demonstrates a reconstruction framework that combines classical PCA with a compact generative AI, establishing a data-driven paradigm for WEPL reconstruction in proton radiography. The primary objective of this study is to demonstrate that combining PCA-driven dimensionality reduction with a compact generative AI network can efficiently and accurately reconstruct WEPL maps from proton radiographs, addressing the limitations of conventional methods as much as possible.

In the field of proton imaging, the terms \emph{water-equivalent path length } and \emph{water-equivalent thickness (WET)} are sometimes used interchangeably to describe the integrated proton stopping power along a trajectory, expressed in centimeters of water. Strictly, WEPL accounts for the proton’s actual (potentially non-straight) path due to scattering, while WET is based on a straight-line path. This makes WEPL slightly greater than WET in heterogeneous media, although the difference is generally small. Throughout this manuscript, we consistently use WEPL. 
In this work, the term “high-dimensional radiograph data” refers to the stack of dose maps obtained using the discrete range-modulation approach.

\section{Materials and methods}\label{section:materialsmethod}
\subsection{Simulation Setup}\label{sub:simulation}
Monte Carlo simulations of energy-resolved proton radiography were conducted using the \texttt{TOPAS} software package~\citep{Perl2012,Polster2015,RamosMendez2015}. To closely simulate clinical image quality, the full geometry and beam data of a clinical \texttt{SUMITOMO} scanning nozzle from Kaohsiung Chang Gung Memorial Hospital Proton Center, as well as a human head CT DICOM dataset from the public \href{https://www.dicomlibrary.com/}{DICOM Library}, were imported into \texttt{TOPAS}. The following physics models were used to account for all relevant secondary particles: \texttt{g4em-standard\_opt4}, \texttt{g4h-phy\_QGSP\_BIC\_HP}, \texttt{g4decay}, \texttt{g4ion-binarycascade}, \texttt{g4h-elastic\_HP}, and \texttt{g4stopping} \citep{novak2019short}.
The DICOM dataset comprised 121 axial CT slices of the human head, each 1.25~mm thick, with a $255 \times 255$~mm$^{2}$ field of view and a $512 \times 512$ image matrix. Within \texttt{TOPAS}, these images were converted to a three-dimensional voxelized head phantom using a modified Schneider algorithm calibrated with clinical CT measurements. The material segmentation assigned up to 25 distinct tissue types with corresponding densities, based on HU values~\citep{schneider2000correlation}.
A pencil beam scanning field of $20 \times 20$~cm$^{2}$ was simulated, with proton energies from 70~MeV to 230~MeV in 2~MeV increments, and $\sim 10^7$ primary protons per energy. The virtual detector was positioned immediately after the phantom, with a sensitive area of $25 \times 25$~cm$^{2}$ and $500 \times 500$ pixel resolution. To obtain projections from multiple directions, the phantom was rotated from $0^\circ$ to $360^\circ$ in $5^\circ$ steps, resulting in 72 projection angles.
For supervised deep learning, ground-truth WEPL maps (in mm) were generated for each projection. Each voxel in the phantom was assigned a stopping power value by multiplying its density with the mass stopping power for that material, calculated using the SRIM database~\citep{Ziegler2010}. For each projection, these values were integrated along the proton beam path using ray-tracing, producing a 2D WEPL map at every angle. These maps served as pixel-wise ground truth for supervised learning. 
\begin{figure}[!htb]
    \centering
    \fbox{%
      \begin{minipage}{0.97\textwidth}
        \centering
        \subfigure[]{\includegraphics[scale=0.3]{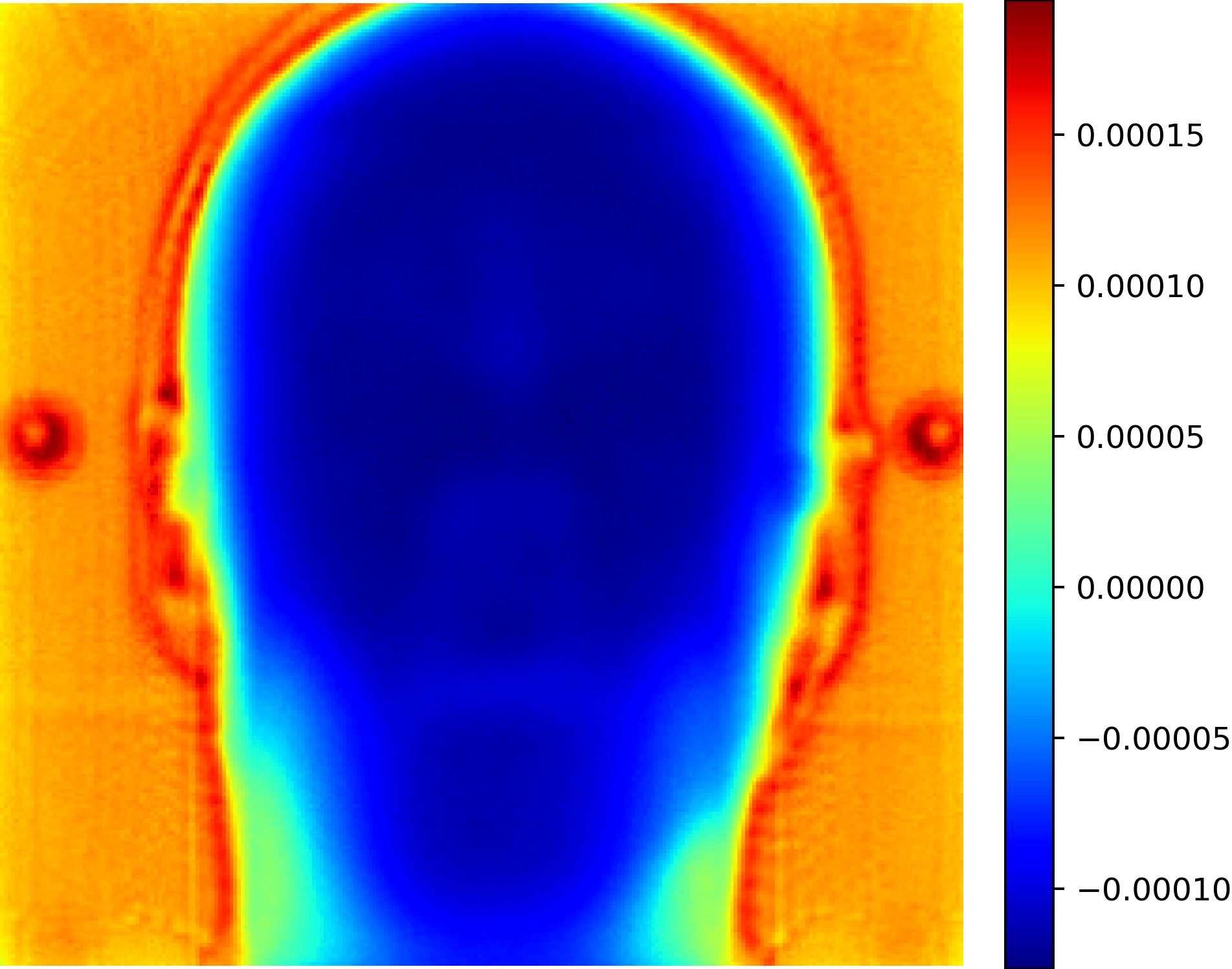} \label{fig:Comp1}}
        \subfigure[]{\includegraphics[scale=0.3]{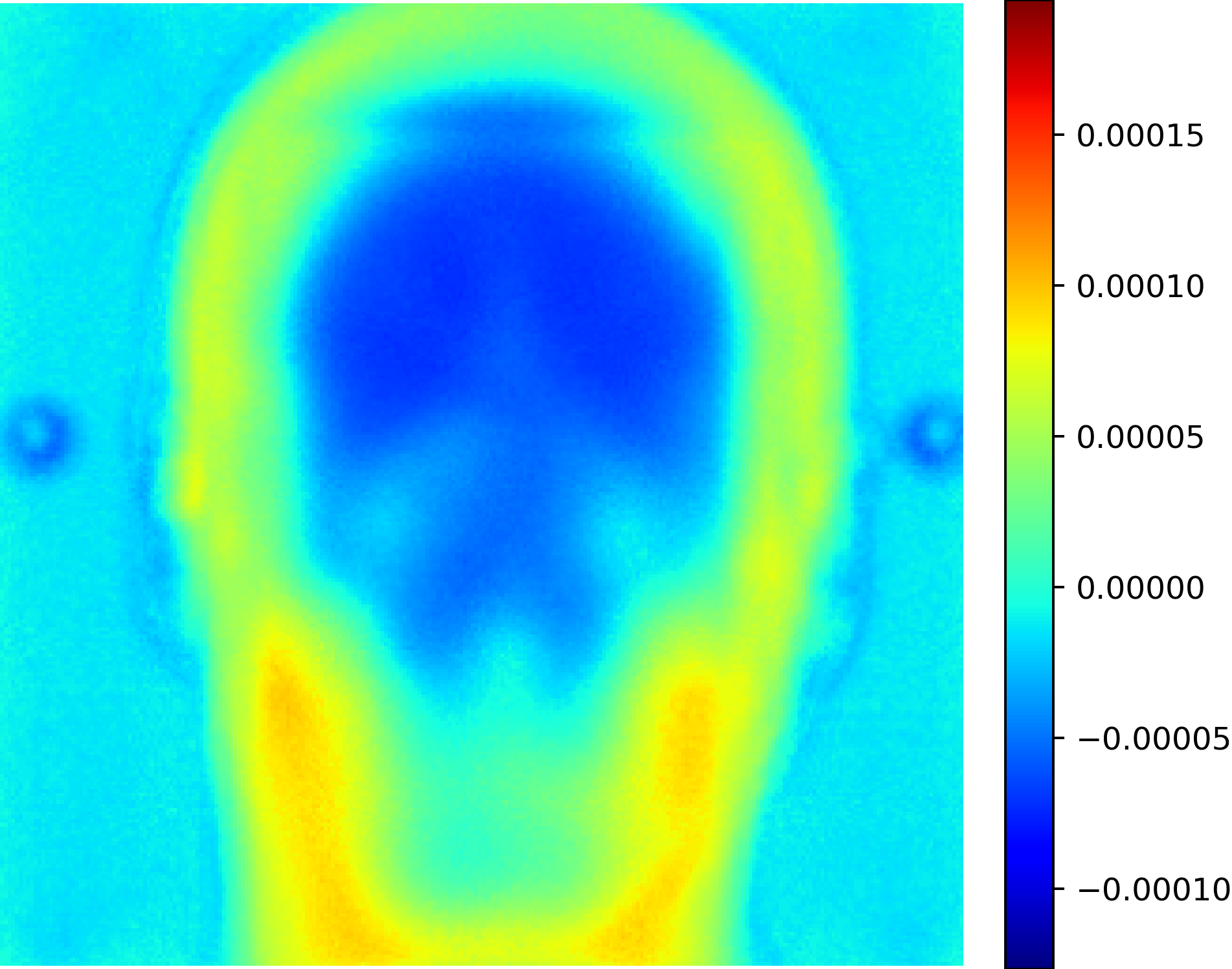} \label{fig:Comp2}} 
        \subfigure[]{\includegraphics[scale=0.3]{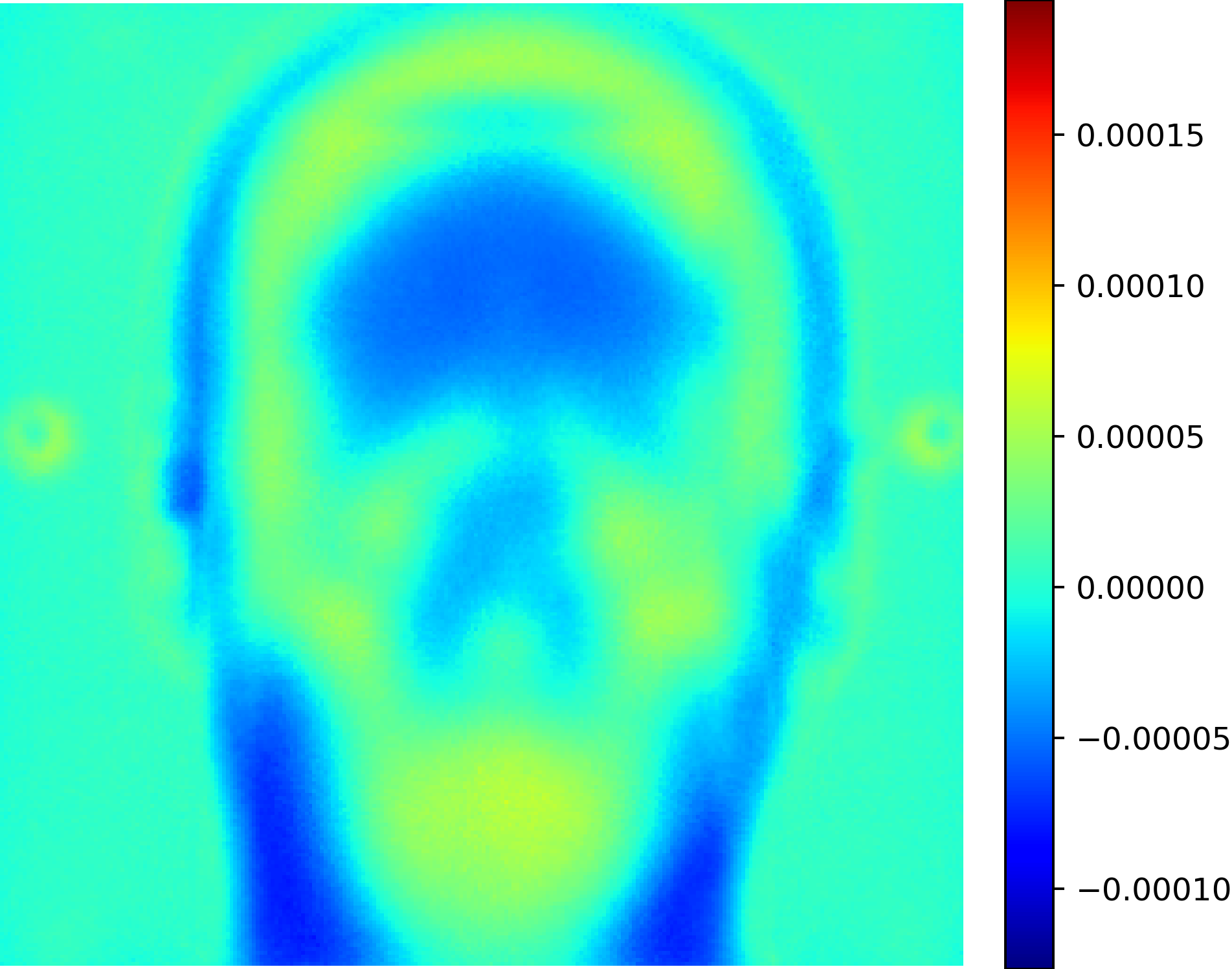} \label{fig:Comp3}} 
        \subfigure[]{\includegraphics[scale=0.3]{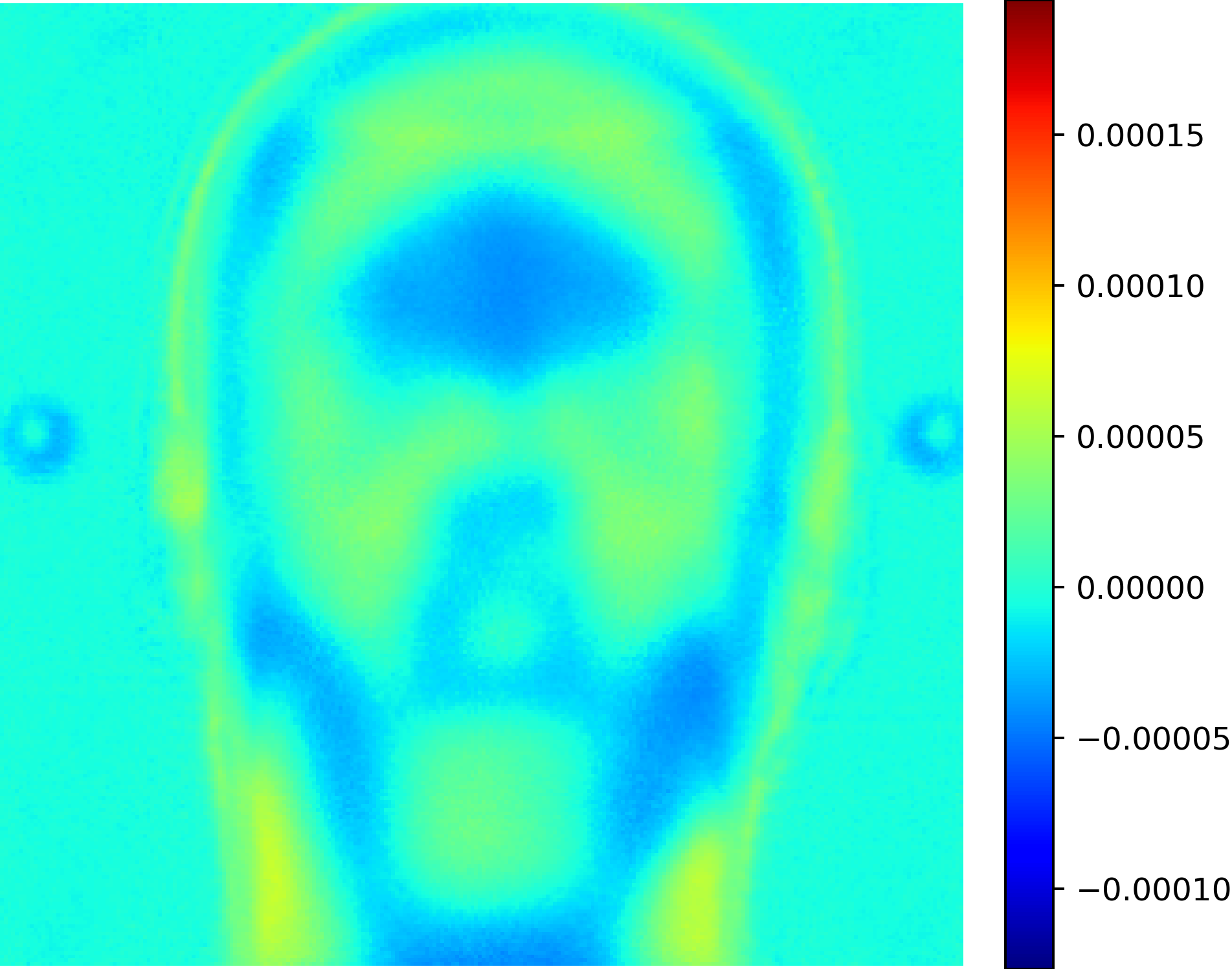} \label{fig:Comp4}}
        \subfigure[]{\includegraphics[scale=0.3]{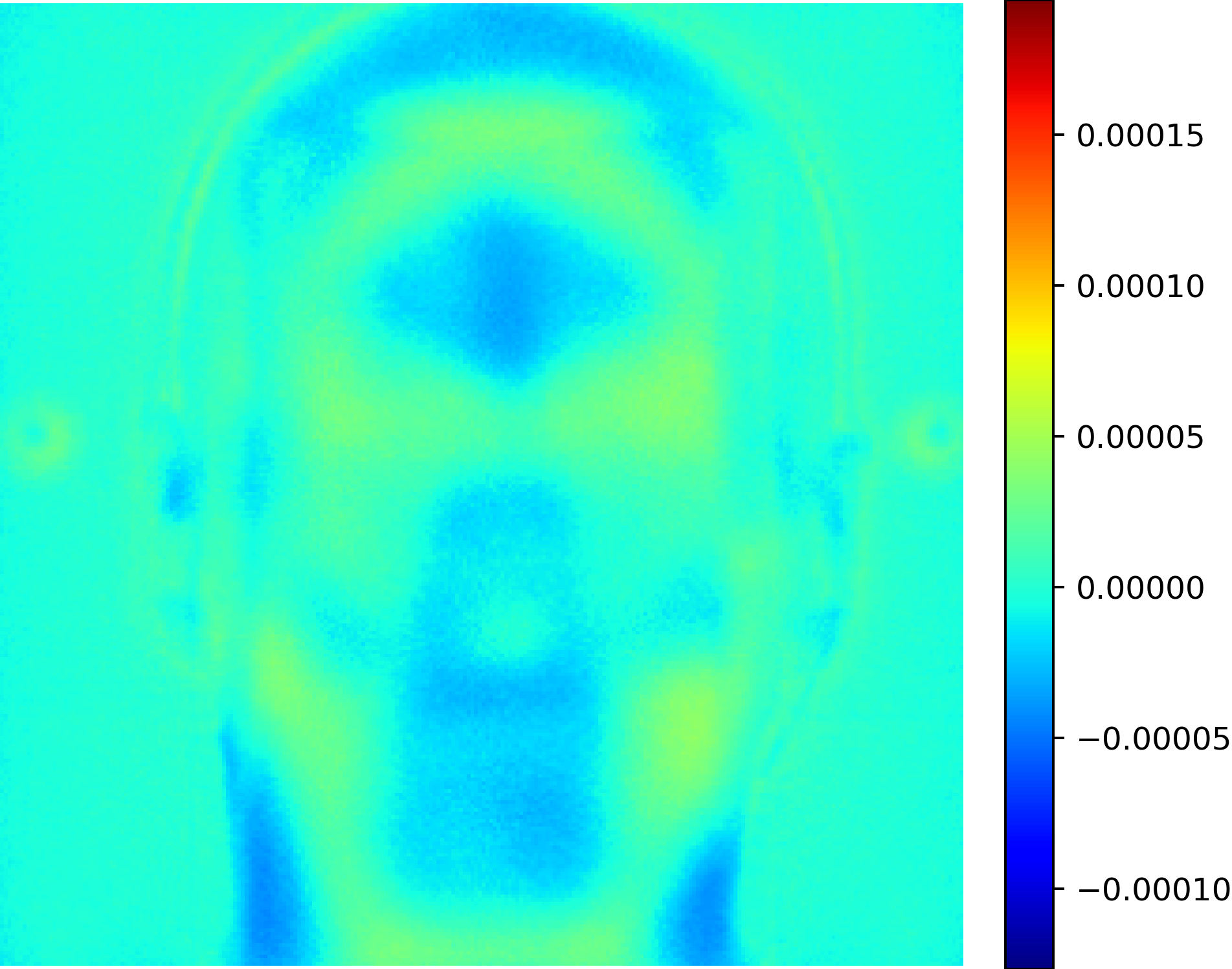} \label{fig:Comp5}} 
        \subfigure[]{\includegraphics[scale=0.3]{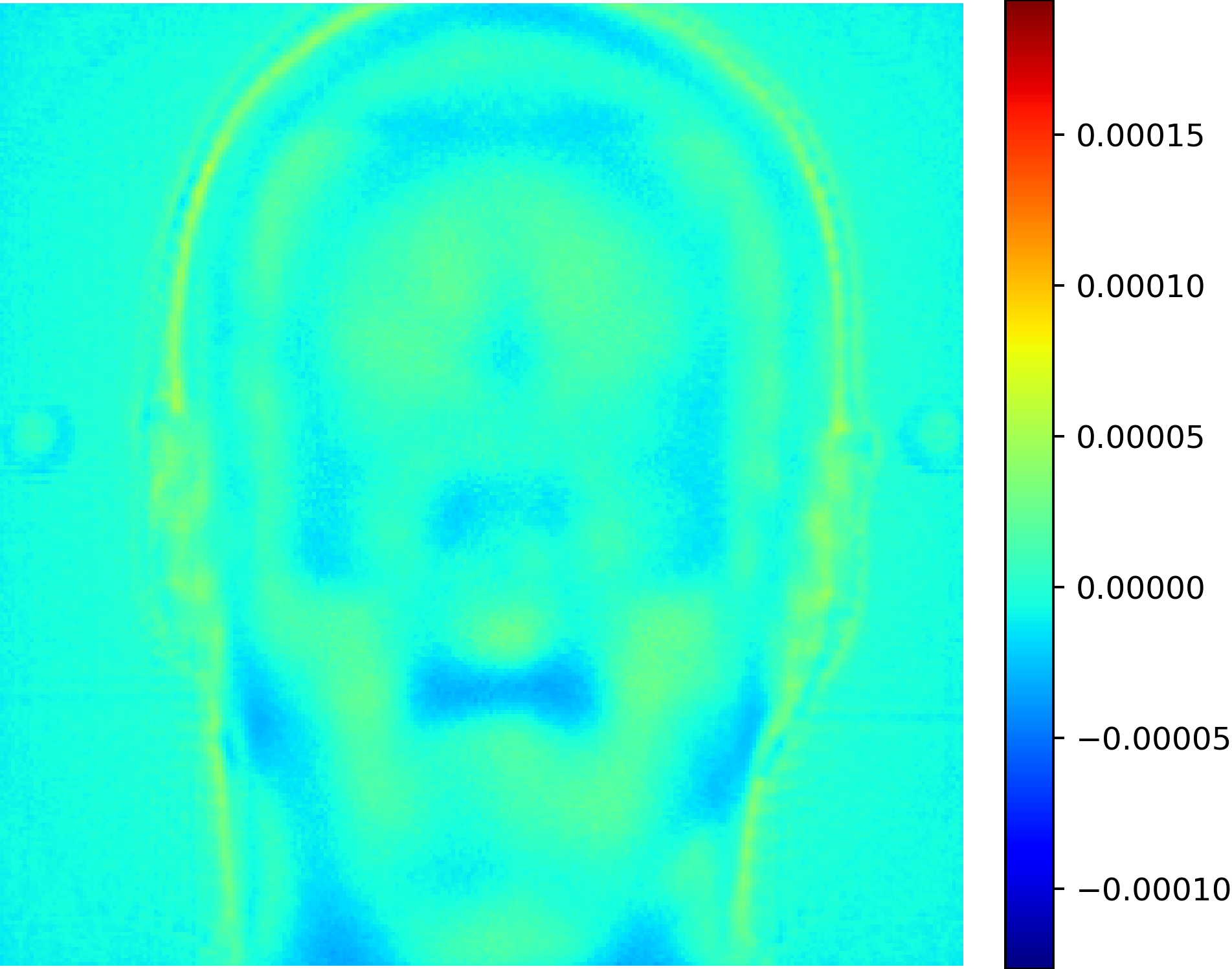} \label{fig:Comp6}} 
        
        \caption{Six linear components extracted using PCA from the high-dimensional radiograph data. Each component captures a distinct mode of variation.}
        \label{fig:linear_components}
      \end{minipage}%
    }
\end{figure}
\subsection{Principal component analysis}\label{sec:dimensionality_reduction}
The discrete range modulation approach in proton radiography produces high-dimensional data, since each detector pixel samples multiple energy channels. This increases computational demands and elevates the risk of overfitting for machine learning models. To mitigate these issues, we used PCA and  implemented with \texttt{scikit-learn}~\citep{pedregosa2011scikit} to reduce data dimensionality, while preserving anatomical detail necessary for accurate WEPL reconstruction.
PCA transforms the mean-centered data matrix $\mathbf{X} \in \mathbb{R}^{n \times d}$ by diagonalizing its covariance matrix $\boldsymbol{\Sigma}$:
\begin{equation}
    \boldsymbol{\Sigma} \mathbf{v}_i = \lambda_i \mathbf{v}_i,
\end{equation}
where $\lambda_i$ are eigenvalues and $\mathbf{v}_i$ the corresponding eigenvectors. The data are projected onto the subspace spanned by the top $k$ eigenvectors:
\begin{equation}
    \mathbf{Z} = (\mathbf{X} - \boldsymbol{\mu}) \mathbf{W}, \qquad \mathbf{W} = [\mathbf{v}_1, \ldots, \mathbf{v}_k],
\end{equation}
with approximate reconstruction given by
\begin{equation}
    \hat{\mathbf{X}} = \mathbf{Z} \mathbf{W}^{\top} + \boldsymbol{\mu},
    \label{eq:inverseRecon}
\end{equation}
where $\boldsymbol{\mu}$ is the mean vector. 

\begin{figure}[!htbp]
    \centering
    \fbox{%
      \begin{minipage}{0.98\textwidth}
        \centering
        \subfigure[]{\label{fig:Comp7a}
            \includegraphics[trim={0cm 0cm 0cm 0cm},clip,scale=0.45]{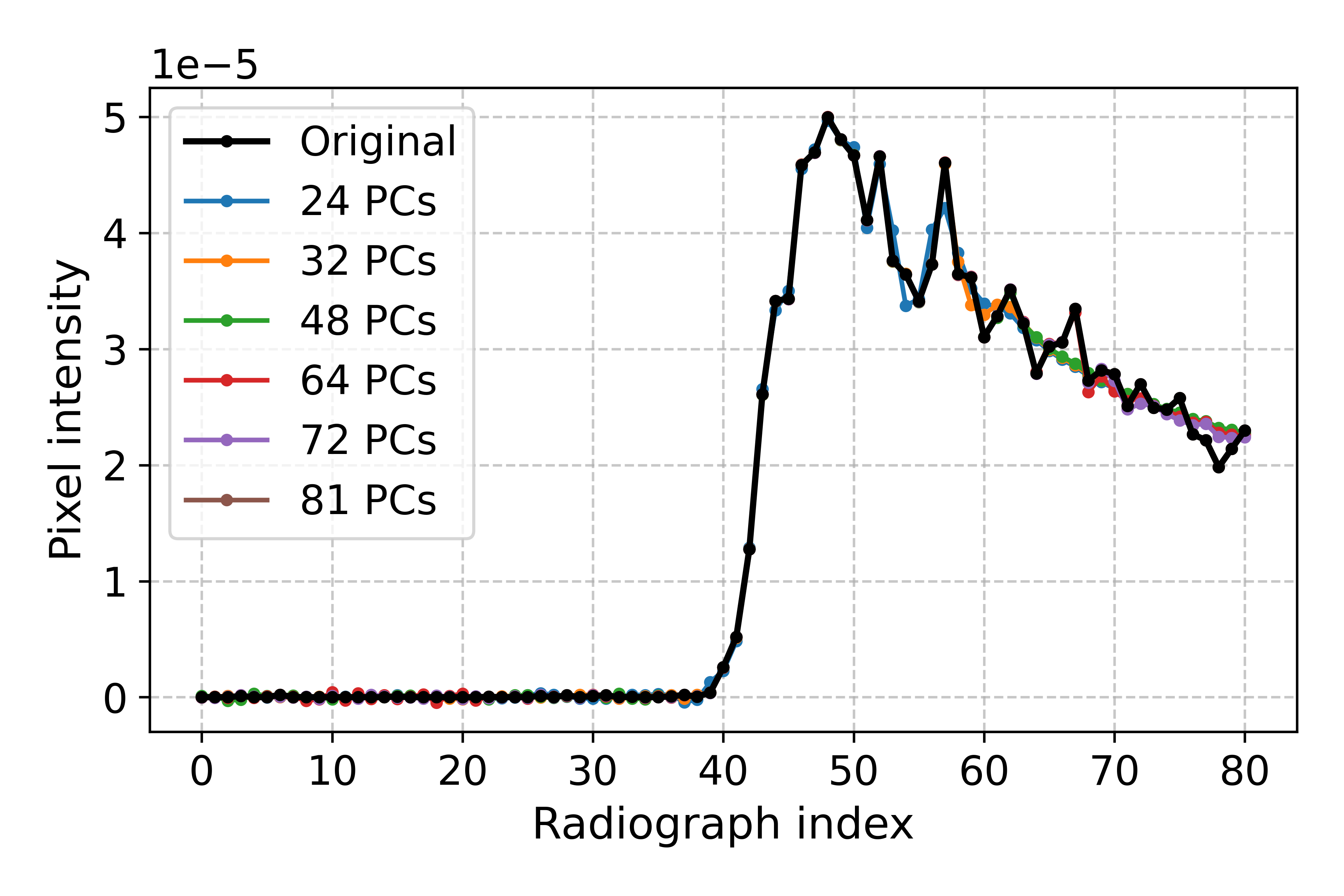}
        }
        \subfigure[]{\label{fig:Comp7b}
            \includegraphics[trim={0cm 0cm 0cm 0cm},clip,scale=0.45]{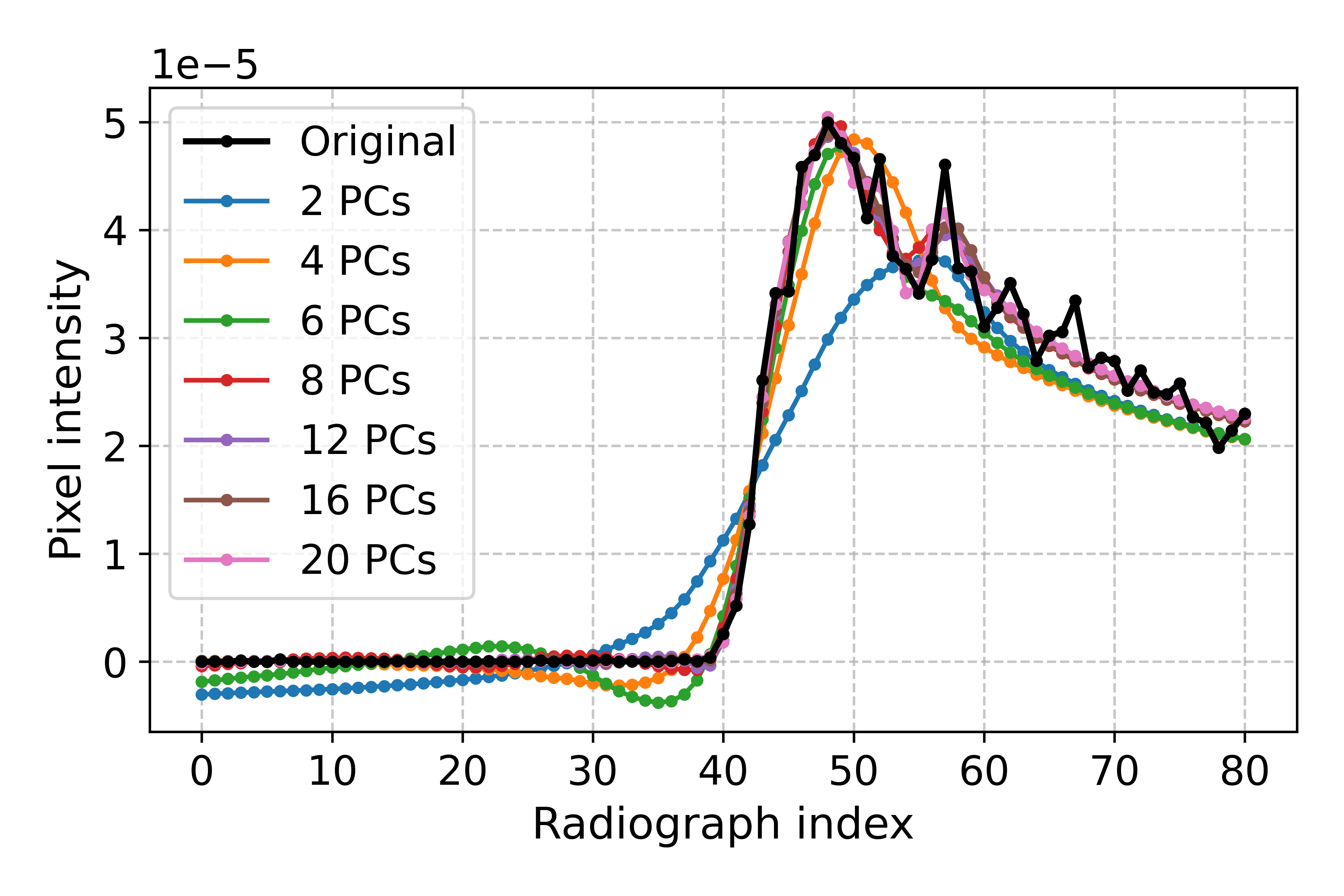}
        }
        \caption{Comparison between the actual central pixel intensity profiles and reconstructed through varying numbers of PCA components using Equation \ref{eq:inverseRecon}. (a) Reconstructed profiles using 24-81 components. (b) Reconstructed profiles using 2-20 components with varying gaps.}
        \label{fig:profilewithcompnets}
      \end{minipage}%
    }
\end{figure}

To balance detail retention with efficiency, we selected $k=16$ principal components which captures approximately 99\% of the data variance as shown in Table~\ref{tab:explained_variance_pca}.  This choice is further supported by improved downstream model performance. Figure~\ref{fig:linear_components} shows the first four principal components, and Figure~\ref{fig:profilewithcompnets} presents central pixel intensity profiles reconstructed with varying component counts using Equation~\ref{eq:inverseRecon} which illustrate the trade-off between reducing dimensionality and retaining anatomical information. Using too many components may reintroduce noise or non-informative variation, while too few may discard essential detail, both of which can degrade WEPL reconstruction accuracy. The PCA components are orthogonal by construction, ensuring independence in captured variance modes. The variance values reported in Table~\ref{tab:explained_variance_pca} were computed across the entire dataset to estimate the mean contribution of each component, providing an intuitive and semi-quantitative  basis for selecting the optimal number of components required for accurate WEPL mapping.
\begin{table}[htbp]
\caption{Mean explained variance for each principal component (PC) across different PCA schemes. The last column reports the cumulative explained variance for all components in each scheme.}
\centering
\begin{adjustbox}{width=\textwidth}
\begin{tabular}{c c c c c c c c c c c c c c c c c c}
\toprule
\textbf{Components} & \textbf{PC1} & \textbf{PC2} & \textbf{PC3} & \textbf{PC4} & \textbf{PC5} & \textbf{PC6} & \textbf{PC7} & \textbf{PC8} & \textbf{PC9} & \textbf{PC10} & \textbf{PC11} & \textbf{PC12} & \textbf{PC13} & \textbf{PC14} & \textbf{PC15} & \textbf{PC16} & \textbf{Total} \\
\midrule
2  & 0.7105 & 0.1490 &  --   &  --   &  --   &  --   &  --   &  --   &  --   &  --   &  --   &  --   &  --   &  --   &  --   &  --   & 0.8594 \\
4  & 0.7105 & 0.1490 & 0.0580 & 0.0274 &  --   &  --   &  --   &  --   &  --   &  --   &  --   &  --   &  --   &  --   &  --   &  --   & 0.9448 \\
6  & 0.7105 & 0.1490 & 0.0580 & 0.0274 & 0.0150 & 0.0100 &  --   &  --   &  --   &  --   &  --   &  --   &  --   &  --   &  --   &  --   & 0.9698 \\
8  & 0.7105 & 0.1490 & 0.0580 & 0.0274 & 0.0150 & 0.0100 & 0.0064 & 0.0047 &  --   &  --   &  --   &  --   &  --   &  --   &  --   &  --   & 0.9808 \\
12 & 0.7105 & 0.1490 & 0.0580 & 0.0274 & 0.0150 & 0.0100 & 0.0064 & 0.0047 & 0.0032 & 0.0022 & 0.0015 & 0.0011 &  --   &  --   &  --   &  --   & 0.9888 \\
16 & 0.7105 & 0.1490 & 0.0580 & 0.0274 & 0.0150 & 0.0100 & 0.0064 & 0.0047 & 0.0032 & 0.0022 & 0.0015 & 0.0011 & 0.0008 & 0.0006 & 0.0005 & 0.0004 & 0.9912 \\
\bottomrule
\end{tabular}
\end{adjustbox}
\label{tab:explained_variance_pca}
\end{table}
\subsection{Deep learning model}\label{sec:deep_learning_model}
Image reconstruction is formulated as learning a transformation
\begin{equation}
    y = T(x),
    \label{eq:transformation}
\end{equation}
where \(x\) is the stack of PCA-reduced proton radiographs (see Section~\ref{sec:dimensionality_reduction}), and \(y\) is the corresponding WEPL map. Here, \(T\) is implemented using a conditional generative adversarial network (cGAN), with the generator \(G\) mapping PCA-compressed radiographs to WEPL maps and the discriminator \(D\) assessing the realism of the generated outputs conditioned on the input.

To promote stable training, we employ a conditional Wasserstein GAN with gradient penalty (WGAN-GP) \citep{wgangp}. The generator and discriminator loss functions~\citep{isola2017image,yang2019multimodal} are:
\begin{align}
    \mathcal{L}_G &= -\mathbb{E}_{x}\left[ D(G(x)\,|\,x) \right], \\
    \mathcal{L}_D &= \mathbb{E}_{y,\,x}\left[D(y\,|\,x)\right] - \mathbb{E}_{x}\left[D(G(x)\,|\,x)\right] \nonumber \\
    &\hspace{2.6cm}+\, \lambda_{\text{gp}}\,\mathbb{E}_{\hat{y}} \left[ \left(\Vert \nabla_{\hat{y}} D(\hat{y}|x) \Vert_2 - 1\right)^2 \right],
\end{align}
where \(x\) is the PCA-reduced input, \(y\) is the real WEPL map, \(\hat{y}\) is an interpolation between real and generated samples, and \(\lambda_{\text{gp}}\) is the gradient penalty weight.

The generator loss comprises an adversarial term (\(\mathcal{L}_G\)), a perceptual loss from VGG16 features, and pixel-level terms (mean squared error (MSE) and structural similarity index (SSIM)):
\begin{equation}
    \mathcal{L}_{\text{generator}} = \mathcal{L}_G + \lambda_{\text{perceptual}} \mathcal{L}_{\text{perceptual}} + \lambda_{\text{combined}} \mathcal{L}_{\text{combined}},
\end{equation}
where \(\mathcal{L}_{\text{combined}}\) combines MSE and SSIM, with weighting parameters \(\lambda_{\text{perceptual}} = 0.1\) and \(\lambda_{\text{combined}} = 50\) set by validation.

\begin{figure}[!htb]
    \centering
    \fbox{%
      \begin{minipage}{1\textwidth}
        \centering
        \subfigure[]{
            \includegraphics[trim={0cm 0cm 0cm 0cm},clip,scale=0.45]{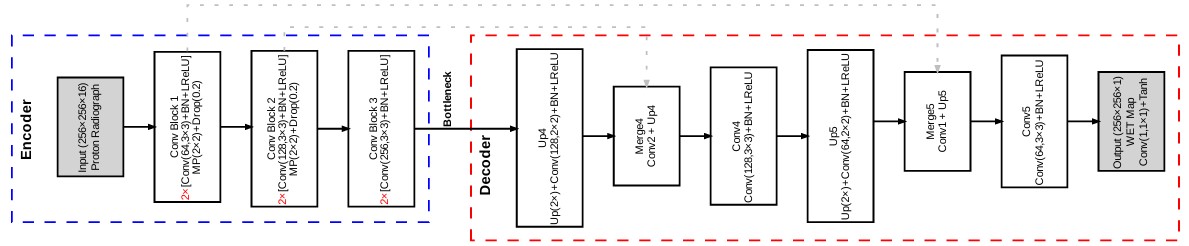}
            \label{fig:unetmodel}
        }
        \subfigure[]{
            \includegraphics[trim={0cm 0cm 0cm 0cm},clip,scale=0.32]{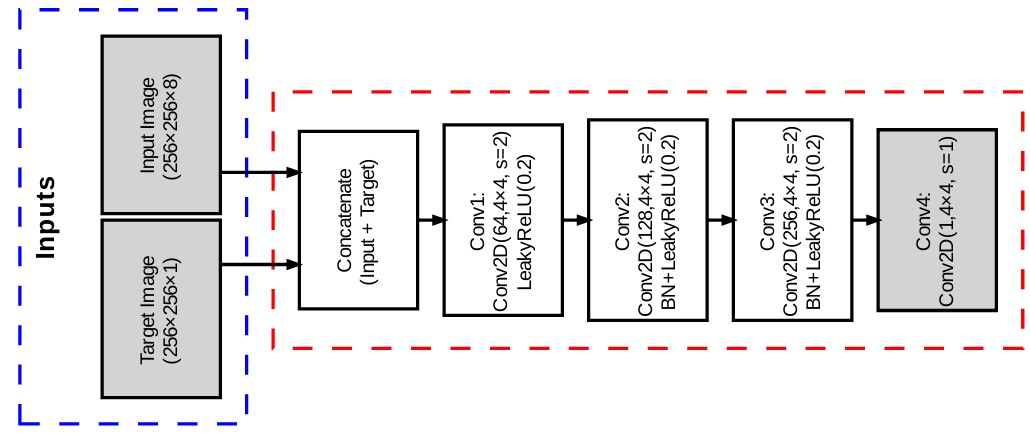}
            \label{fig:Discr}
        }
        \subfigure[]{
            \includegraphics[trim={0cm 0cm 0cm 0cm},clip,scale=0.35]{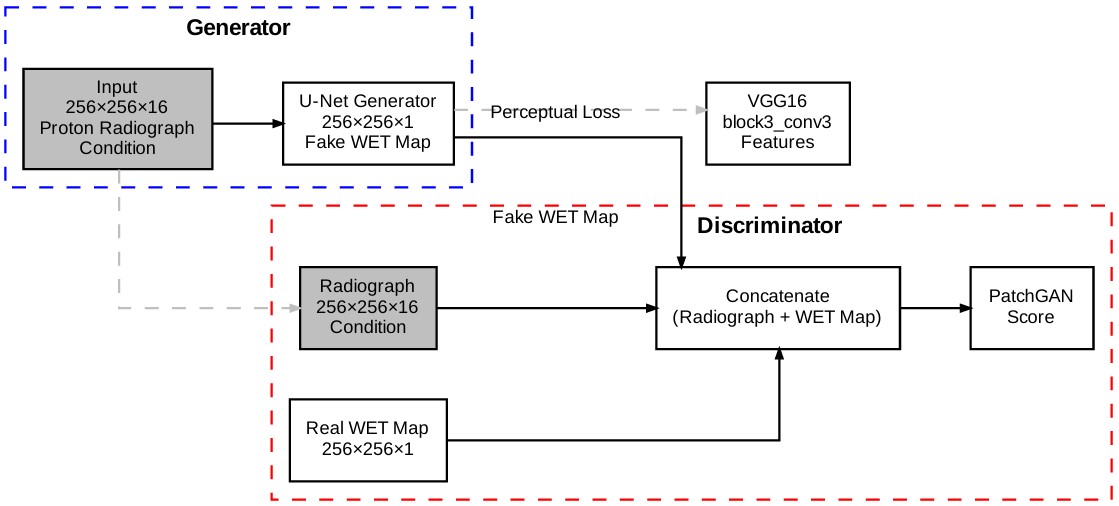}
            \label{fig:GAN}
        }
       \caption{
Overview of the cGAN pipeline.  
(a) The U-Net generator converts proton radiographs \mbox{(256$\times$256$\times$16)} to WEPL maps \mbox{(256$\times$256$\times$1)} via an encoder--decoder with skip connections.  
(b) The patch discriminator assesses the realism of concatenated radiograph--WEPL map pairs at the patch level.  
(c) The cGAN framework combines generator and discriminator for end-to-end, high-fidelity WEPL map prediction.
} \label{fig:UNETGANDic}
\end{minipage}
    }
\end{figure}

Figure~\ref{fig:UNETGANDic} illustrates the cGAN architecture, showing the U-Net generator and PatchGAN discriminator. The model uses PCA-compressed radiograph stacks as conditional input, rather than a noise vector unlike conventional image generation approaches.
The VGG16 perceptual loss promotes the preservation of anatomical structures and spatial detail in the reconstructed WEPL maps, complementing pixel-level losses and reducing the risk of overly smooth or unrealistic outputs.

\subsection{Data preparation}\label{sub:data_preparation}

Each sample in this study comprises a stack of 81 discrete range-modulated radiograph images, acquired at different proton energies, and one corresponding WEPL map, as described in Section~\ref{sub:simulation}. The original image size for both the radiographs and WEPL maps was $250 \times 250$ pixels. As a preprocessing step, all images were resampled to $256 \times 256$ pixels using bilinear interpolation with anti-aliasing, ensuring compatibility with conventional deep learning architectures that often require kernel sizes and feature map dimensions to be multiples of four~\citep{pedregosa2011scikit}.

The dataset was constructed from CT images of four different patients, yielding a total of 277 samples. To ensure reliable model evaluation and minimize bias, the data were partitioned into three sets: 64\% for training (177 samples), 10\% for validation (28 samples), and 26\% for testing (72 samples). This was achieved using a test split of 0.2 to define the test set, followed by a validation split of 0.1 from the remaining data. A random seed of 42 was used for reproducibility~\citep{dutta2022seed}. Data splits were stratified by patient to prevent data leakage, ensuring that samples from a single patient were not present in multiple subsets.

Normalization parameters (minimum and maximum values) were computed exclusively on the training set and subsequently applied to normalize all data to the range $[-1, 1]$, thereby improving numerical stability during training. Radiograph stacks were transposed to the shape (samples, height, width, channels), while WEPL maps were expanded with a channel dimension for compatibility with the model input requirements.

The validation set was used to monitor performance and guide hyperparameter tuning during training mentioned in Table~(\ref{tab:gen_hyperparams}-\ref{tab:disc_hyperparams}), whereas the test set was reserved strictly for final evaluation. This partitioning strategy maximizes the utility of all available data and provides an unbiased assessment of the model's generalization capability to unseen cases.
\begin{subsection}{Model architecture, training, and implementation}\label{sub:model_training}
We implemented a conditional generative adversarial network  to map multi-channel proton radiographs, with 16 channels per sample after PCA, to WEPL maps. The generator employs a U-Net \citep{ronneberger2015u} encoder--decoder architecture for efficient spatial feature extraction, as illustrated in Figure~\ref{fig:unetmodel}. This design is effective for medical image translation tasks requiring precise boundary preservation. Detailed generator hyperparameters are provided in Table~\ref{tab:gen_hyperparams}. The discriminator uses a PatchGAN design~\citep{demir2018patch}, which evaluates local image patches for realism by concatenating input and output images, as shown in Figure~\ref{fig:Discr}. Discriminator hyperparameters are summarized in Table~\ref{tab:disc_hyperparams}.

The training loss function combines a Wasserstein adversarial loss with gradient penalty, MSE, SSIM, and perceptual loss based on VGG16 features. The loss weights, tuned on the validation set, are detailed in Tables~\ref{tab:gen_hyperparams} and \ref{tab:disc_hyperparams}. Training used the Adam optimizer with a learning rate of $2 \times 10^{-4}$, $\beta_1 = 0.5$, batch size of 8, and up to 9000 epochs. Early stopping, based on validation loss with a patience of 200 epochs, and model checkpointing were applied to prevent overfitting.

All experiments were conducted on a workstation with an Intel Core Ultra 9 285K CPU (24 cores), 64~GB RAM, and an NVIDIA RTX 2080 Ti GPU (11~GB VRAM) \citep{nvidia2020ampere}, running Ubuntu 24.04, Python 3.11, and TensorFlow 2.18.0 with CUDA 12.5.1 and cuDNN 9. The final cGAN model, including both U-Net generator and PatchGAN discriminator, comprises approximately 15 million trainable parameters, enabling efficient end-to-end training without computational or memory limitations.
\end{subsection}

\begin{table}[!htb]
\caption{Generator (U-Net) hyperparameters.}
\label{tab:gen_hyperparams}
\centering
\begin{adjustbox}{width=0.62\textwidth}
\begin{tabular}{lc}
\hline
\textbf{Parameter} & \textbf{Value} \\
\hline
Architecture & U-Net, encoder--decoder with skip connections \\
Input channels & 16 (post-PCA) \\
Filters per block & 64 / 128 / 256 \\
Kernel size & $3 \times 3$ \\
Activation & LeakyReLU ($\alpha=0.1$) \\
Batch normalization & Yes (all conv layers) \\
Max pooling size & $2 \times 2$ \\
Dropout rate & 0.2 (after pooling) \\
Regularization & $l_1=1 \times 10^{-5}$, $l_2=1 \times 10^{-4}$ \\
Output activation & \texttt{tanh} ($1 \times 1$ conv) \\
Loss weights & $\lambda_{\text{perceptual}} = 0.1$, $\lambda_{\text{combined}} = 50$ \\
\hline
\end{tabular}
\end{adjustbox}
\end{table}

\begin{table}[!htb]
\caption{Discriminator (PatchGAN) hyperparameters.}
\label{tab:disc_hyperparams}
\centering
\begin{adjustbox}{max width=0.62\textwidth}
\begin{tabular}{lc}
\hline
\textbf{Parameter} & \textbf{Value} \\
\hline
Architecture & PatchGAN (input/output concatenation) \\
Filters per block & 64 / 128 / 256 \\
Kernel size & $4 \times 4$ \\
Activation & LeakyReLU ($\alpha=0.2$) \\
Batch normalization & Yes (2nd, 3rd conv layers) \\
Output layer & $1 \times 1$ conv, patch score map \\
Loss & Wasserstein + gradient penalty ($\lambda_{\text{gp}} = 5$) \\
\hline
\end{tabular}
\end{adjustbox}
\end{table}

\subsection{Evaluation metrics}\label{subsec:metric}
Model performance was assessed by comparing the predicted WEPL maps to the ground truth using five quantitative metrics: mean absolute error (MAE), root mean squared error (RMSE), structural similarity index (SSIM), Kullback–Leibler (KL) divergence, and the proton radiography gamma passing rate.

MAE quantifies the average absolute error per pixel:
\begin{equation}
\text{MAE} = \frac{1}{N} \sum_{i=1}^N \left| y^{\mathrm{true}}_i - y^{\mathrm{pred}}_i \right|,
\end{equation}
where $y^{\mathrm{true}}_i$ and $y^{\mathrm{pred}}_i$ denote the true and predicted WEPL (in mm) for pixel $i$, and $N$ is the total number of pixels in the test set.

RMSE penalizes larger errors and is given by:
\begin{equation}
\text{RMSE} = \sqrt{ \frac{1}{N} \sum_{i=1}^N \left( y^{\mathrm{true}}_i - y^{\mathrm{pred}}_i \right)^2 }.
\end{equation}

SSIM measures perceptual and structural similarity between the predicted and reference WEPL maps~\citep{SSIM2012}:
\begin{equation}
\text{SSIM}(x, y) = \frac{(2\mu_x \mu_y + C_1)(2\sigma_{xy} + C_2)}{(\mu_x^2 + \mu_y^2 + C_1)(\sigma_x^2 + \sigma_y^2 + C_2)},
\end{equation}
where $\mu_x, \mu_y$ are the mean WEPL values, $\sigma_x^2, \sigma_y^2$ are variances, and $\sigma_{xy}$ is the covariance of images where $x$  represents the ground truth and $y$  represents the prediction.

KL divergence quantifies the similarity of value distributions between the ground truth and predicted WEPL maps~\citep{belov2011}:
\begin{equation}
D_{\mathrm{KL}}(P \parallel Q) = \sum_{i} P(i) \log\left( \frac{P(i)}{Q(i)} \right),
\end{equation}
where $P(i)$ and $Q(i)$ are the normalized histograms (probability mass functions) of the ground truth and predicted WEPL maps, respectively.

Quantitative assessment of proton radiography (pRG) image quality is essential for benchmarking different reconstruction methods. However, no  clinically relevant standard metric currently exists for this purpose. To address this gap, we use  the gamma index for proton radiography ($\gamma_{pRG}$) index is a novel metric also used in \citep{Tsai2022DRM}. This $\gamma_{pRG}$ inspired by the gamma index~\citep{low2010gamma,Saito2022GammaIndexJapan}, which is widely used in radiotherapy for spatial and dosimetric comparison of dose distributions. While the conventional gamma index evaluates agreement between dose distributions, the $\gamma_{pRG}$ is specifically adapted to compare WEPL maps by incorporating both spatial proximity and WEPL differences:

\begin{equation}
\gamma_\mathrm{pRG}(i, j) = \min_{(x, y)} \sqrt{
    \left( \frac{T(x, y) - R(i, j)}{\Delta \mathrm{WEPL}} \right)^2 +
    \left( \frac{x - i}{\Delta d} \right)^2 +
    \left( \frac{y - j}{\Delta d} \right)^2
}
\label{eq:gammapRG}
\end{equation}

Here, $R(i, j)$ and $T(x, y)$ denote the ground truth and predicted WEPL values, respectively, at the indicated pixel locations; $\Delta \mathrm{WEPL}$ represents the allowable WEPL tolerance (set to 2\% of the local value), and $\Delta d$ is the spatial tolerance (set to 3\,mm, corresponding to 4 pixels in our images). For each reference pixel, the minimum $\gamma_\mathrm{pRG}$ value within a defined search neighborhood is calculated. The $\gamma_{pRG}$ \textit{passing rate} is then defined as the percentage of pixels with $\gamma_\mathrm{pRG} < 1$, analogous to the passing rate interpretation in standard gamma analysis. Passing rates above 95\% indicate excellent agreement between predicted and reference WEPL maps.

\section{Results}\label{section:Results}

\subsection{Prediction of water-equivalent path length}\label{subsec:WEPL_prediction}

\begin{figure}[!htb]
    \centering
    \fbox{%
      \begin{minipage}{1\textwidth}
        \centering
        \subfigure[]{
            \includegraphics[trim={9cm 0cm 8cm 0cm},clip,scale=0.22]{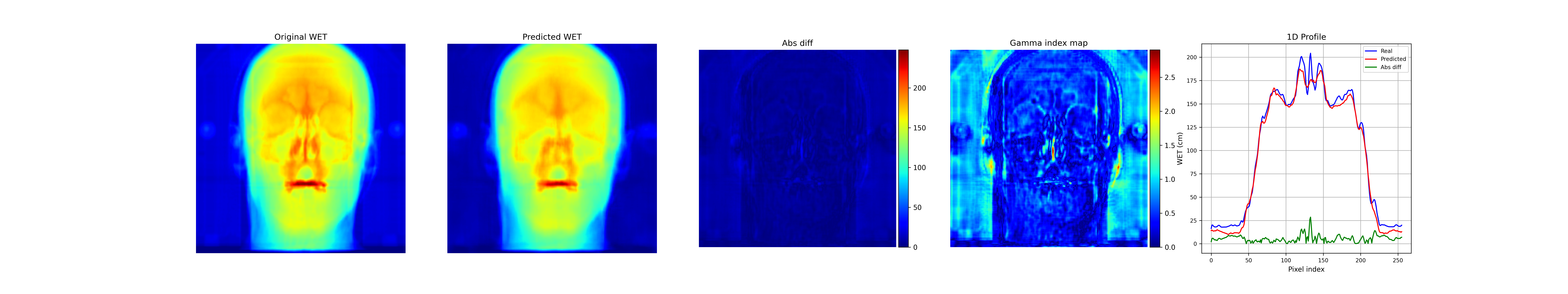}
            \label{fig:comparison_wetSamples0}
        }
        \subfigure[]{
            \includegraphics[trim={9.cm 0cm 8cm 0cm},clip,scale=0.22]{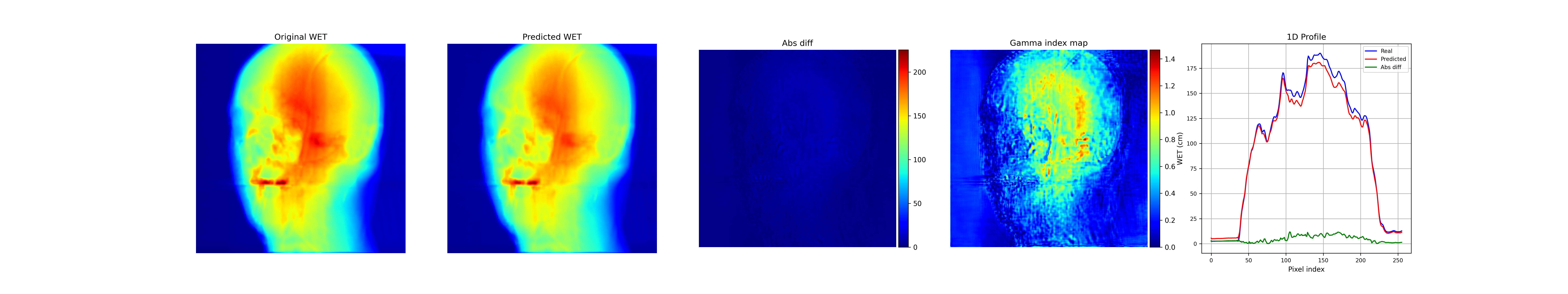}
            \label{fig:comparison_wetSamples29}
        }
        \caption{
            Comparison of predicted and ground truth WEPL maps with gamma index. 
            From left to right: (1) Original WEPL (mm) image used as reference. (2) Predicted WEPL (mm) map reconstructed from the model. (3) Absolute difference between prediction and reference. (4) 2D gamma index map computed using spatial and WEPL tolerance.
            (5) Corresponding 1D profile across the center three rows of the image, showing real, predicted, and absolute difference WEPL values.
        }
        \label{fig:comparison_wetSamples}
      \end{minipage}%
    }
\end{figure}

The predictive accuracy of the proposed cGAN-based model for WEPL mapping was assessed visually and quantitatively. Figure~\ref{fig:comparison_wetSamples} illustrates two representative test cases, each comprising a set of five images: the original WEPL map (mm) as ground truth, the predicted WEPL map reconstructed by the model, the absolute difference map highlighting voxel-wise errors, the 2D gamma index map computed with spatial tolerance $\Delta d = 3$~mm and WEPL tolerance $\Delta \mathrm{WEPL} = 2\%$, and the 1D profile across the center rows comparing real, predicted, and absolute difference values. The results demonstrate that predicted WEPL maps are closely aligned with the reference images, with the most prominent errors localized at sharp anatomical boundaries. The gamma index analysis shows that the majority of image pixels satisfy the $\gamma < 1$ criterion, consistent with clinical quality assurance standards for spatial and quantitative agreement. The 1D profiles further confirm that, within the main anatomical regions, the predicted WEPL distributions are nearly indistinguishable from ground truth, and absolute errors remain low throughout the images.

\subsection{Performance evaluation}

\begin{figure}[!htbp]
    \centering
    \fbox{%
      \begin{minipage}{0.96\textwidth}
        \centering
        \subfigure[]{\label{fig:mae1}
            \includegraphics[trim={0cm 0cm 0cm 0cm},clip,scale=.33]{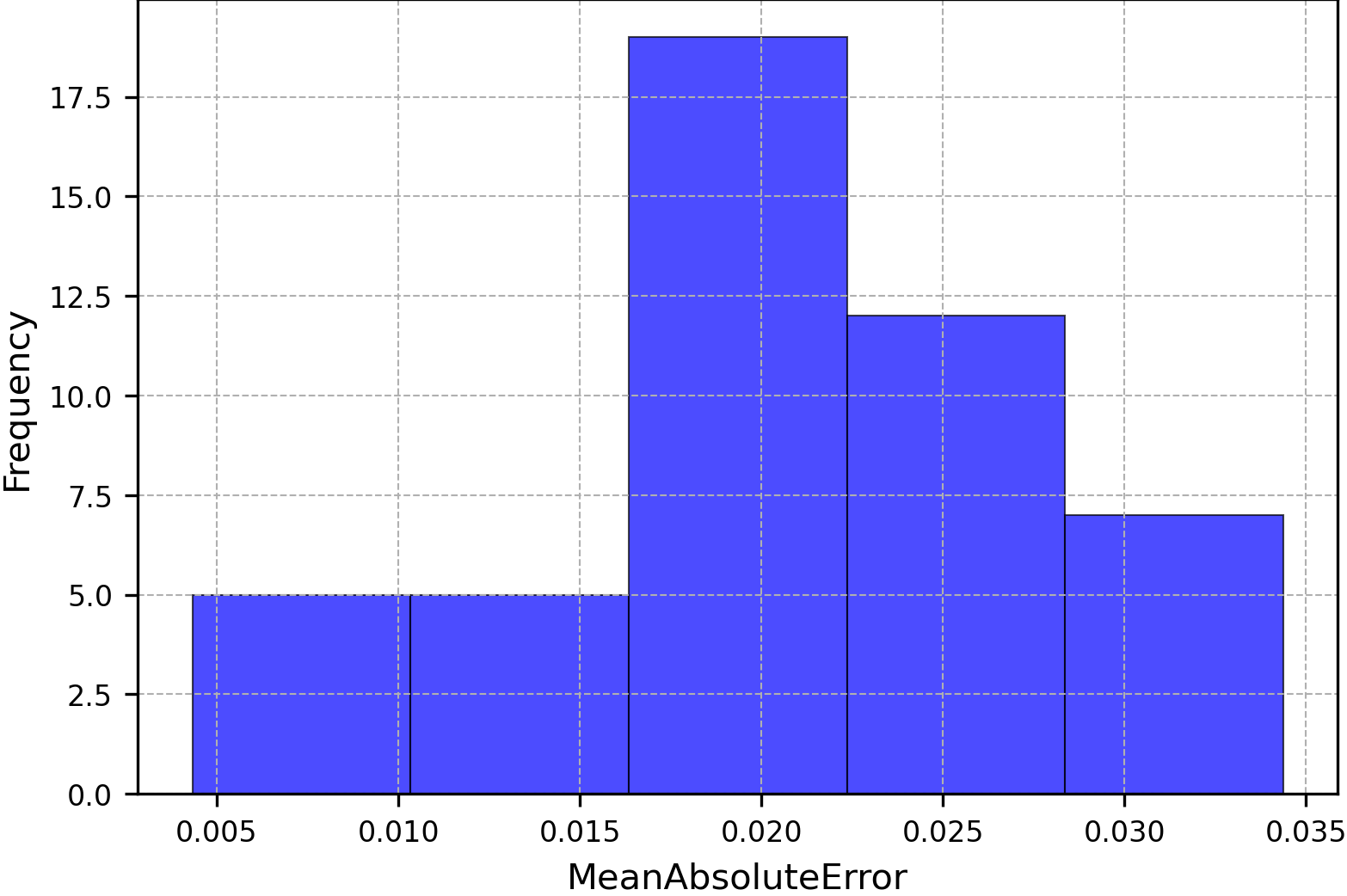}}
        \subfigure[]{\label{fig:rmse1}
            \includegraphics[trim={0cm 0cm 0cm 0cm},clip,scale=.33]{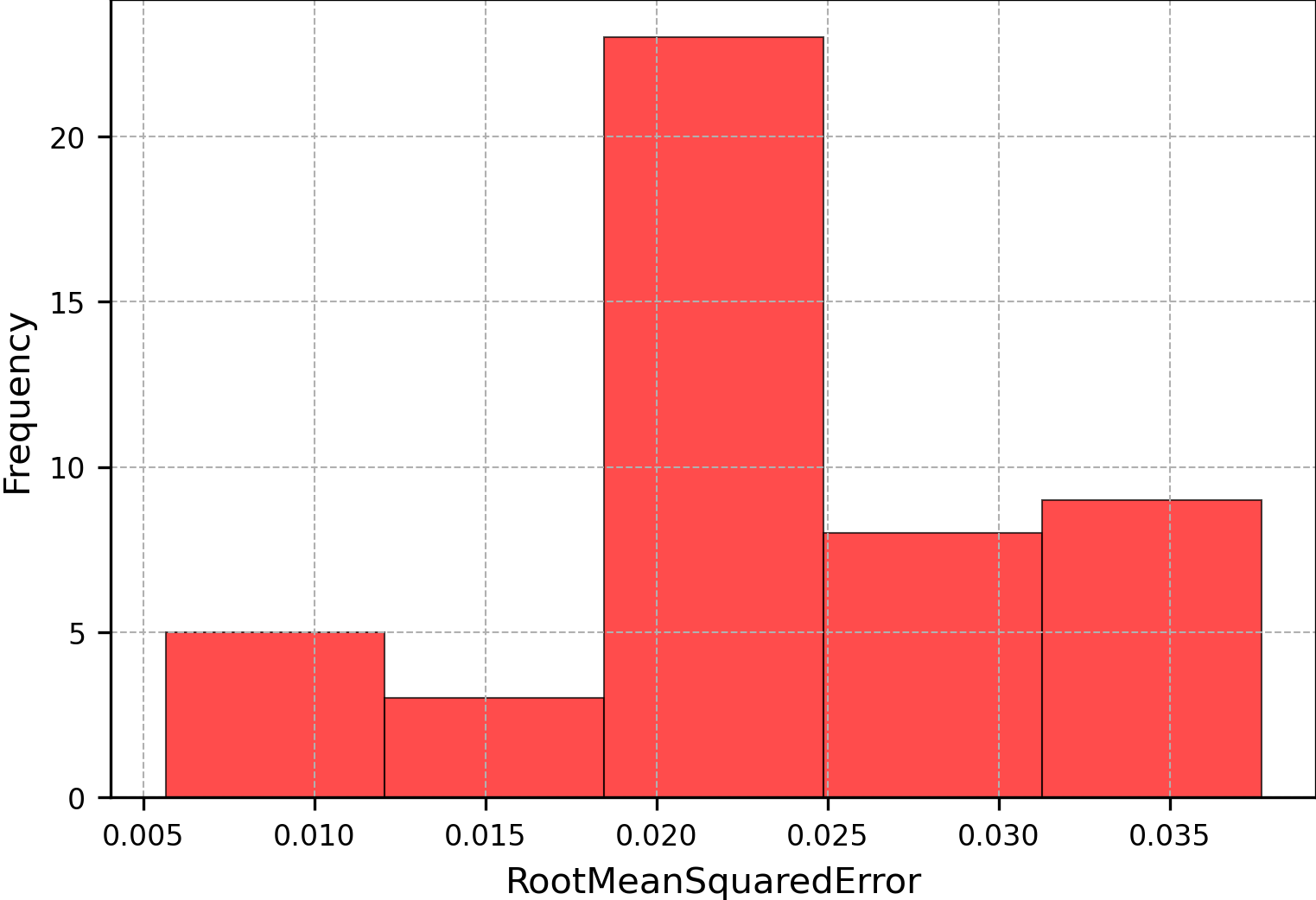}}
        \subfigure[]{\label{fig:ssim1}
            \includegraphics[trim={0cm 0cm 0cm 0cm},clip,scale=.33]{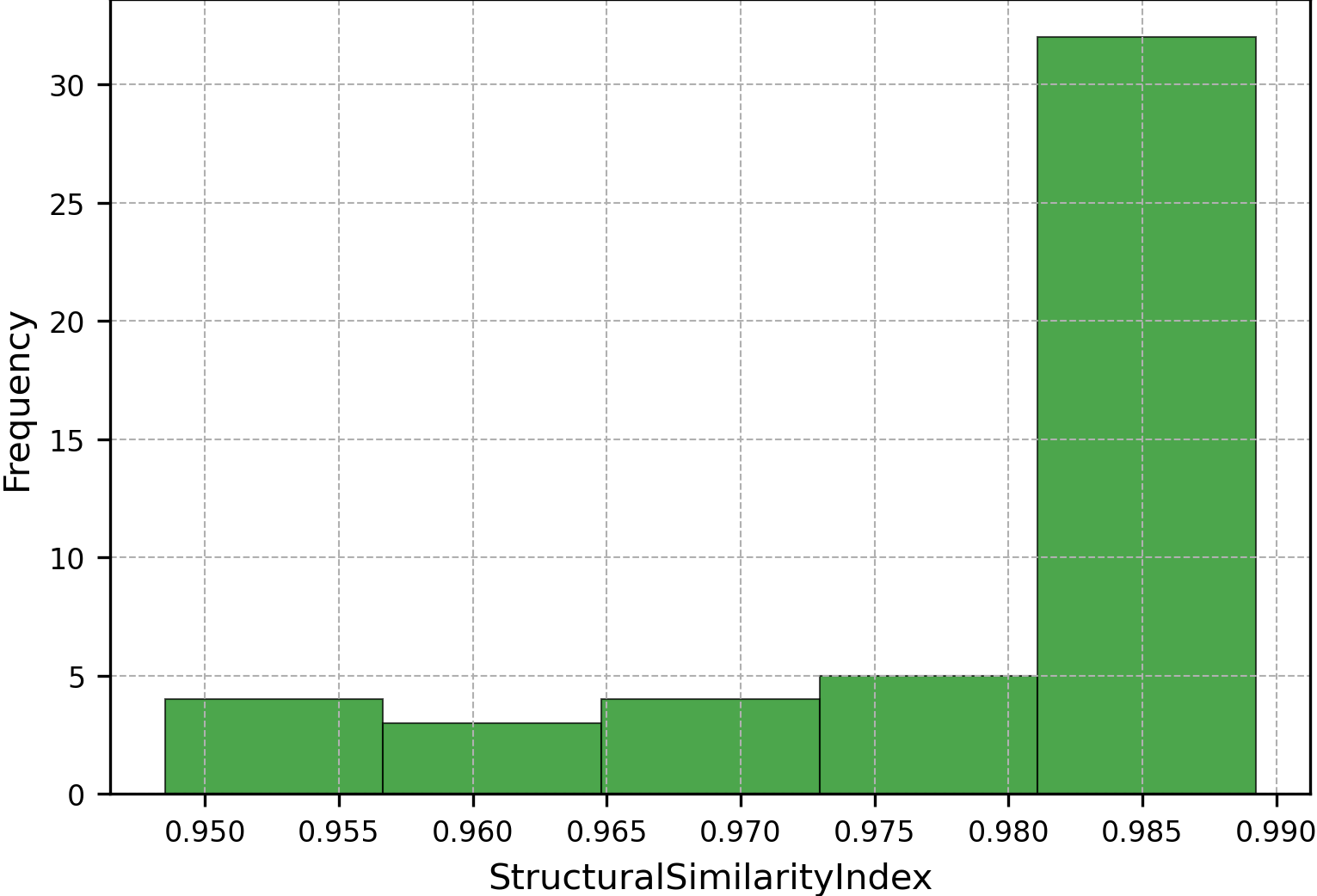}}
        \subfigure[]{\label{fig:kl1}
            \includegraphics[trim={0cm 0cm 0cm 0cm},clip,scale=.33]{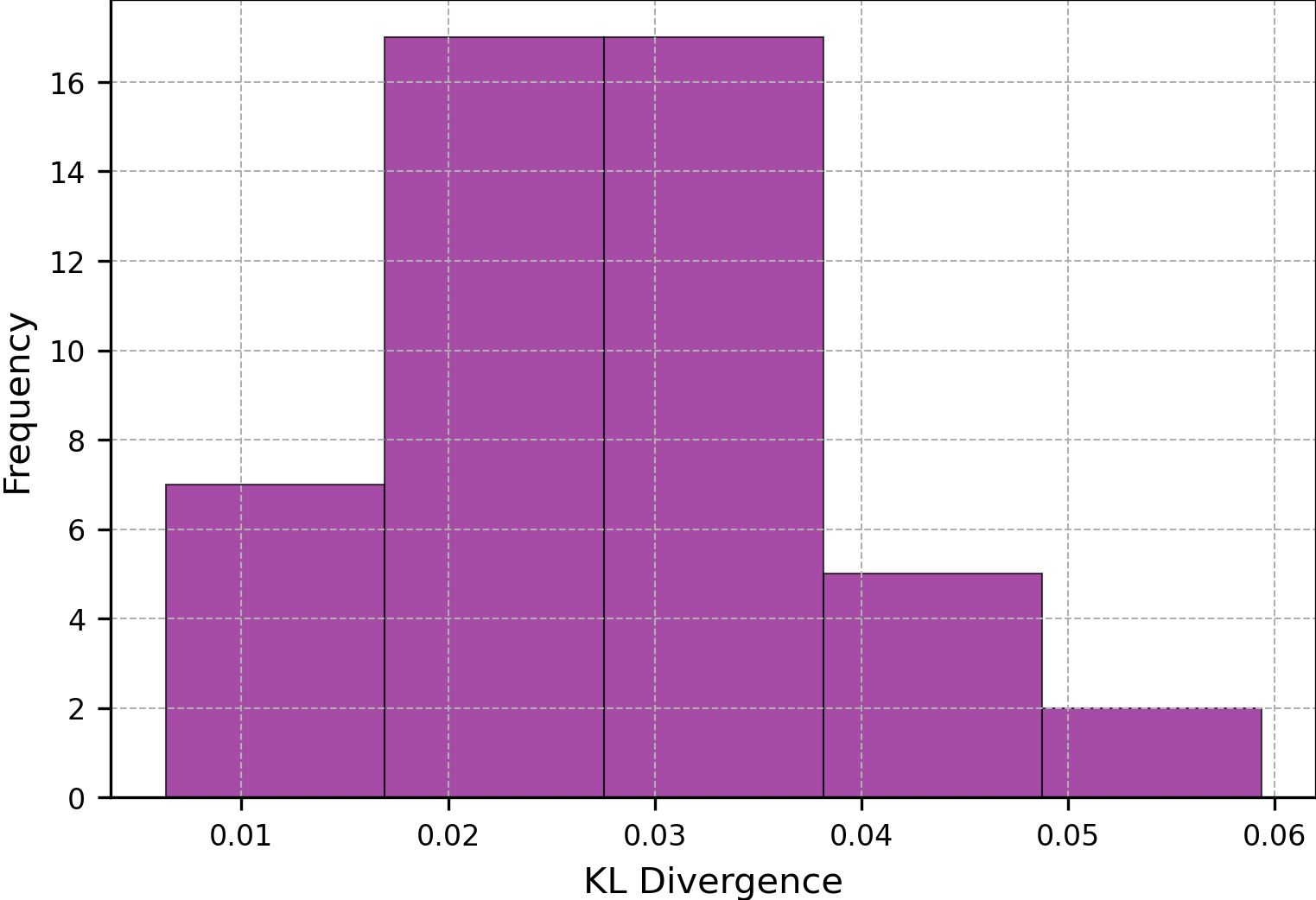}}
        \subfigure[]{\label{fig:gI}
            \includegraphics[trim={0cm .08cm 0cm 0cm},clip,scale=.420]{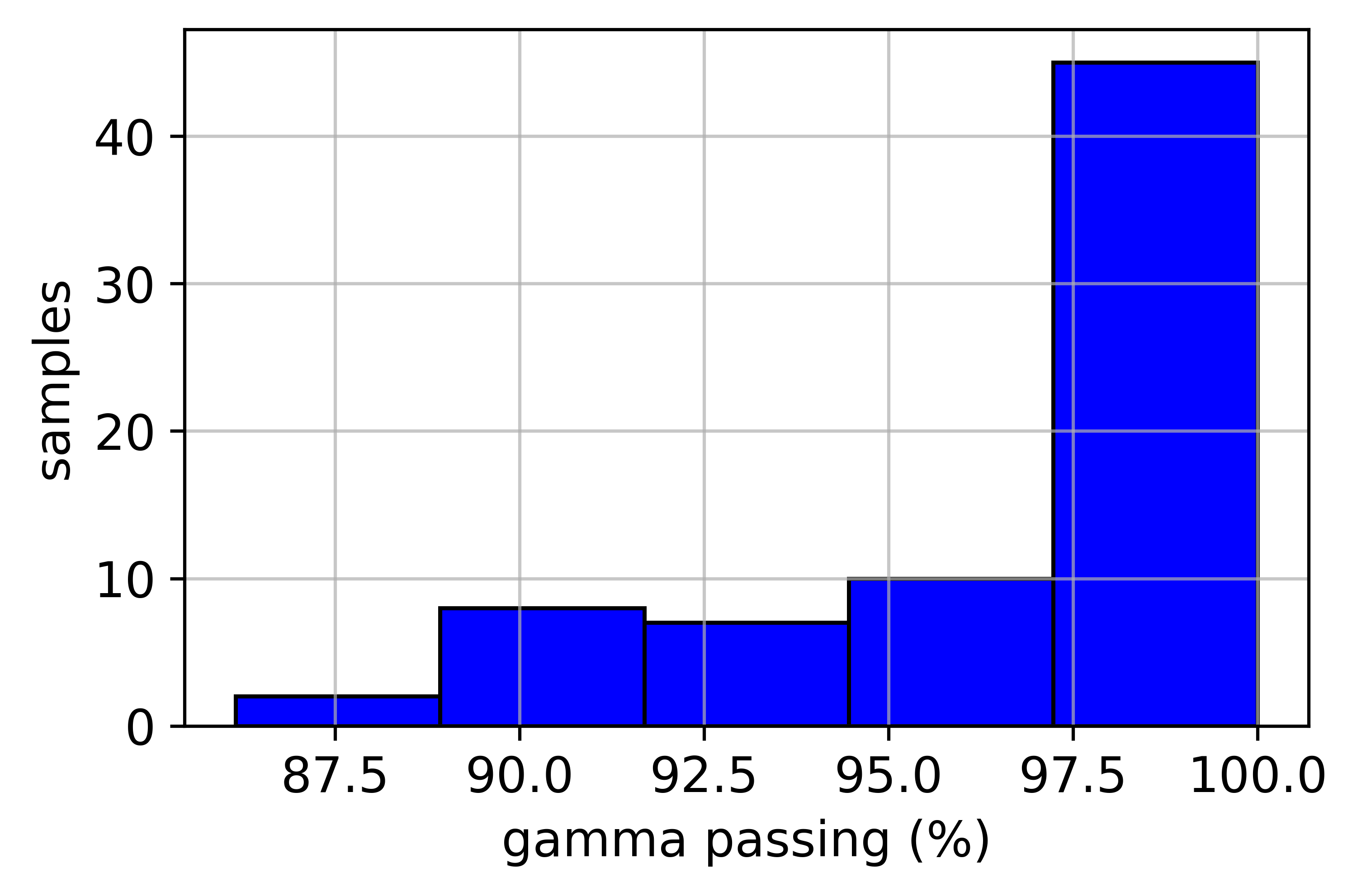}}
        
        \caption{Histograms of the evaluation metrics across test samples. 
                 (a) Mean absolute error.
                 (b) Root mean squared error.
                 (c) Structural similarity index.
                 (d) KL divergence.
                 (e) Gamma passing rate.}
        \label{fig:five_metrics_histograms}
      \end{minipage}%
    }
\end{figure}

Model performance was quantitatively assessed across the test cohort using the metrics described in Section~\ref{subsec:metric}. All metrics were computed for each individual test sample, using the model weights from the final training snapshot where loss values had stabilized and the model had converged. For each sample, we evaluated MAE, RMSE, SSIM, KL divergence, and  $\gamma_{pRG}$. 
The distribution of these metric values across all test samples is shown as histograms in Figure~\ref{fig:five_metrics_histograms}. Panel (a) presents the distribution of MAE, (b) shows RMSE, (c) displays SSIM, (d) shows KL divergence, and (e) illustrates the gamma passing rate. These histograms demonstrate that the majority of test samples have low error values and high structural similarity, with most gamma passing rates clustering at the high end, indicating strong agreement between predicted and reference WEPL maps.
The mean and standard deviation for each metric, calculated across all test samples, are summarized in Table~\ref{tab:evaluation_metrics}. The mean values reflect the central tendency of model performance, while the standard deviation for each metric, also shown as the width of the histograms, serves as an estimate of variability and is used as the error bar. Together, these results indicate that the proposed model achieves consistently high accuracy and robust performance across the entire test cohort.

\begin{table}[!htbp]
  \caption{Evaluation metrics on test datasets. Values are mean $\pm$ standard deviation.}
  \label{tab:evaluation_metrics}
  \centering
  \setlength{\tabcolsep}{5pt}
  \renewcommand{\arraystretch}{0.5}
  \begin{tabular}{lccccc}
    \br
   PCA  & MAE & RMSE & SSIM & KL Div.& $\gamma_{pRG}$ passing (\%) \\
    \br
    16 & 0.025 $\pm$ 0.0072 & 0.028 $\pm$ 0.0075 & 0.971 $\pm$ 0.0206 & 0.020 $\pm$ 0.0069 & 97.0 $\pm$ 3.77 \\
    \br
  \end{tabular}
\end{table}

\subsection{Epoch-wise evolution of metrics} \label{sec:epoch_evolution}

\begin{figure}[!htbp]
    \centering
    \fbox{%
        \begin{minipage}{0.8\textwidth}
            \centering
            \includegraphics[width=0.9\textwidth]{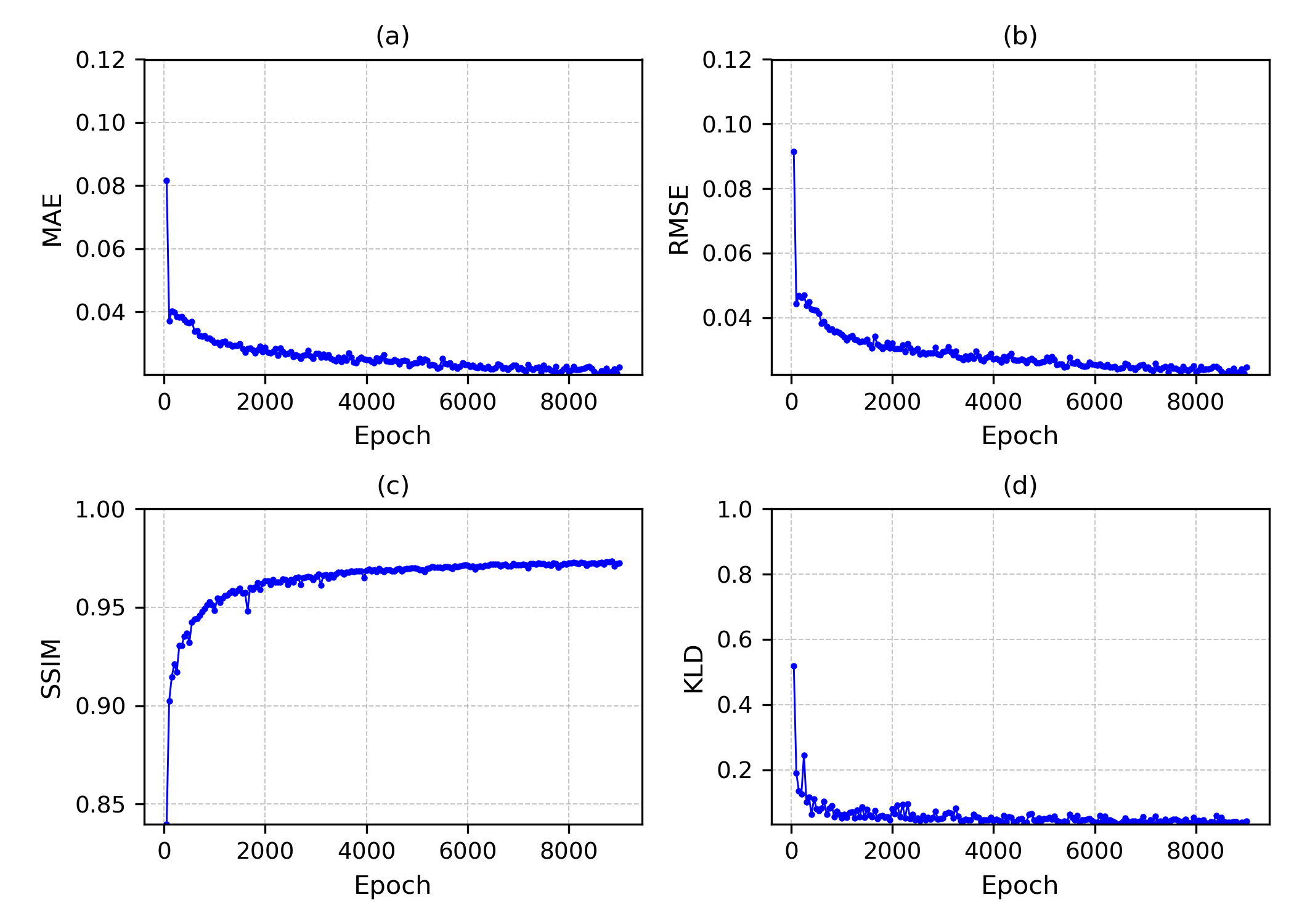}
            \caption{Epoch-wise evolution of four evaluation metrics during model training.
                (a) MAE. (b) RMSE. (c) SSIM. (d) KLD.}
            \label{fig:epochvsmetrics}
        \end{minipage}
    }
\end{figure}

The progression of the four primary evaluation metrics, MAE, RMSE, SSIM, and KL divergence, throughout model training is summarized in Figure~\ref{fig:epochvsmetrics}. At epoch 0, MAE and RMSE begin at their maximum values, approximately 0.11 and 0.12, while SSIM starts near 0.86 and KL divergence at 0.21. The model exhibits rapid learning during the initial 500 epochs, where both MAE and RMSE decrease by more than 50\%, SSIM increases above 0.95, and KL divergence drops to about 0.12. 

After this initial phase, the rate of improvement slows, with all metrics continuing to converge over subsequent epochs. By epoch 3,000, MAE and RMSE stabilize around 0.04, while SSIM reaches a plateau near 0.98, and KL divergence settles at approximately 0.11. This non-linear evolution demonstrates the model’s ability to quickly capture key features in the early stages, followed by diminishing returns as it refines its predictions. 

Beyond 5,000 epochs, the values of all metrics remain stable, indicating that the model has converged and no overfitting or instability is observed. Figure~\ref{fig:normalized_loss} shows the normalized loss component trajectories during training. Each loss term, including adversarial, MSE, SSIM, and perceptual losses, rapidly decreases and then stabilizes. The perceptual and gradient penalty losses show slightly more fluctuation, but no single loss dominates at any point, confirming balanced optimization.

\subsection{Loss function evolution during training}

\begin{figure}[!htbp]
    \centering
    \includegraphics[width=0.55\textwidth]{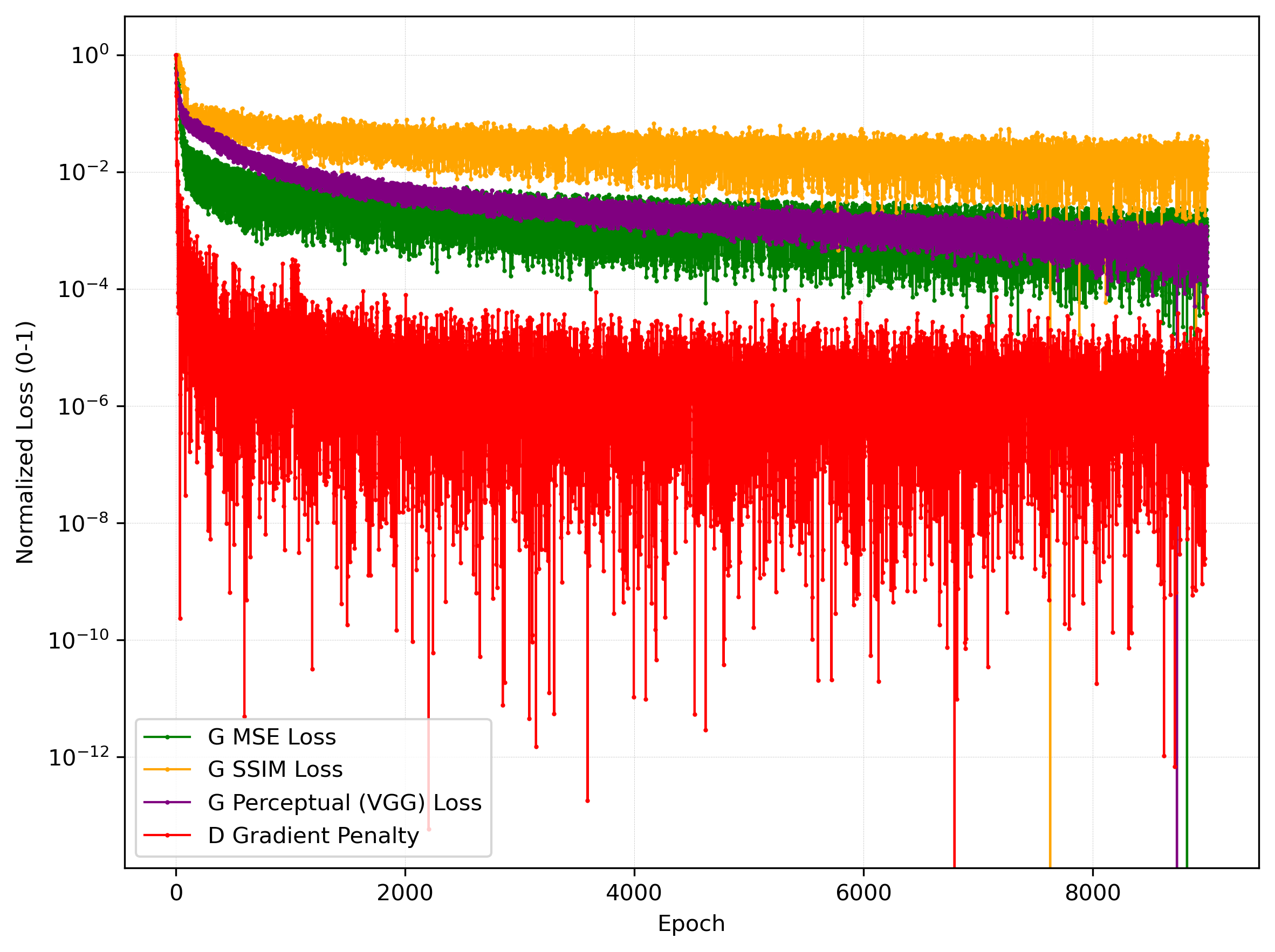}
    \caption{Epoch-wise evolution of individual loss components during training, normalized to the interval $[0,~1]$.}
    \label{fig:normalized_loss}
\end{figure}

Figure~\ref{fig:normalized_loss} shows the epoch-wise evolution of the main loss components during model training. To enable direct comparison across different loss terms, each curve is normalized to the interval $[0,~1]$ based on its initial value, and plotted on a logarithmic scale. This normalization allows visualization of both rapid early decreases and slower, long-term convergence trends in all loss components.

The plot tracks four primary losses: MSE loss, SSIM loss, perceptual  loss for the generator, and the discriminator gradient penalty. At epoch~0, all loss values are set to one. During the initial 500 epochs, MSE and SSIM losses drop by more than 50\%, reflecting rapid improvement in pixel-level and structural similarity between predicted and ground-truth WEPL maps. For both MSE and SSIM, the loss values reach approximately half of their starting value within the first 200--300 epochs, and continue to decline, stabilizing near their baseline by around 3000 epochs. The VGG perceptual loss decreases at a similar rate but exhibits larger fluctuations due to the sensitivity of high-level feature extraction. The gradient penalty loss associated with the discriminator drops most steeply, spanning over eight orders of magnitude within the first few hundred epochs, and then remains at a low level, indicating the  stable adversarial dynamics.

Beyond 3000 epochs, all loss components show slow, incremental improvement and eventually plateau. By epoch 5000, the losses have converged to steady values, with only minor oscillations persisting for the remainder of training. No loss component displays instability or divergence at any point. The combined convergence of all terms demonstrates balanced multi-objective optimization, and supports the consistent high performance of the trained model in WEPL map prediction.

\subsection{Noise sensitivity analysis}\label{NoiseSensitivity}

\begin{figure}[!htbp]
    \centering
    \fbox{%
        \begin{minipage}{0.7\linewidth}
            \centering
            \subfigure[]{\includegraphics[width=0.4\linewidth]{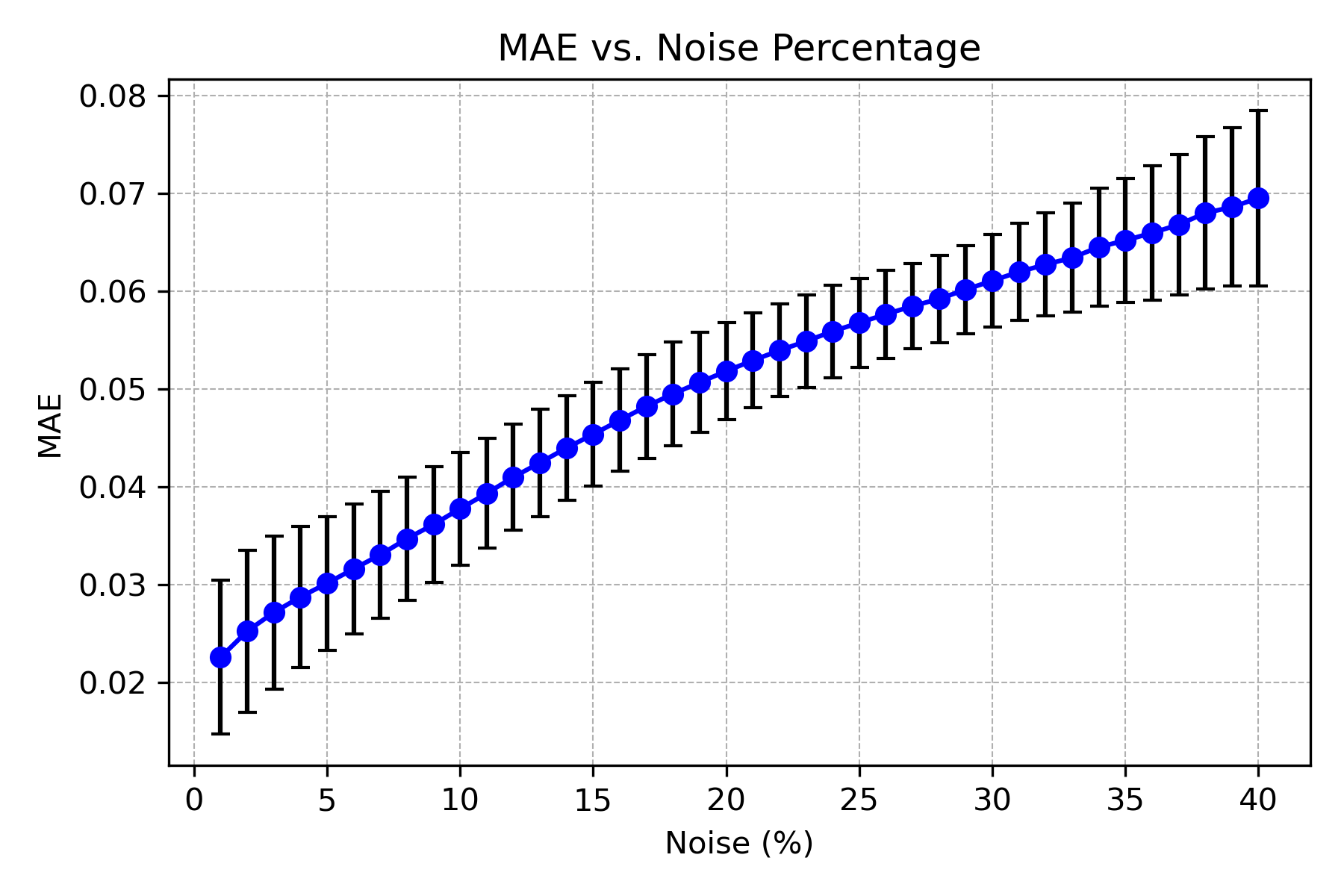}\label{fig:mae}}
            \subfigure[]{\includegraphics[width=0.4\linewidth]{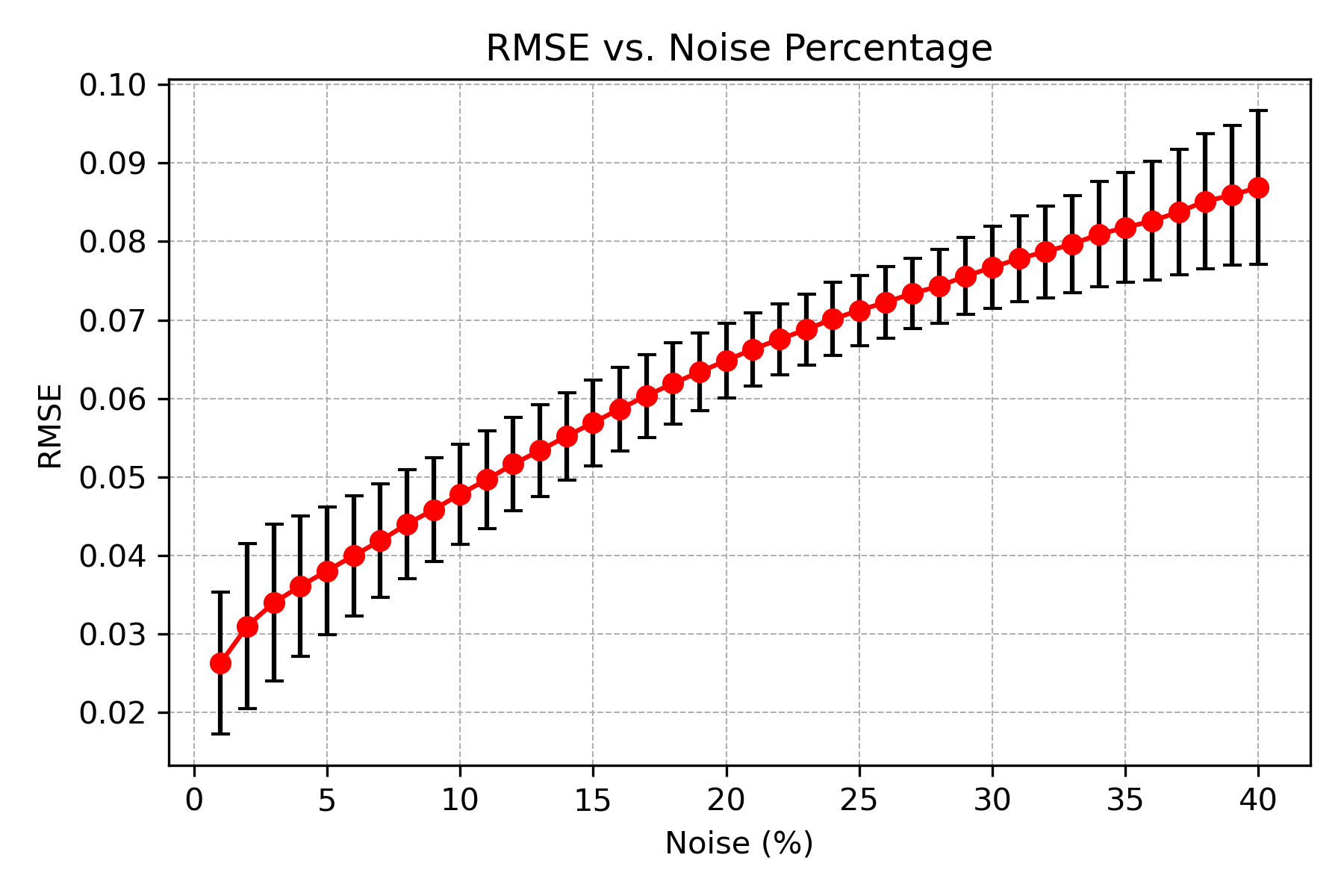}\label{fig:rmse}}
            \subfigure[]{\includegraphics[width=0.4\linewidth]{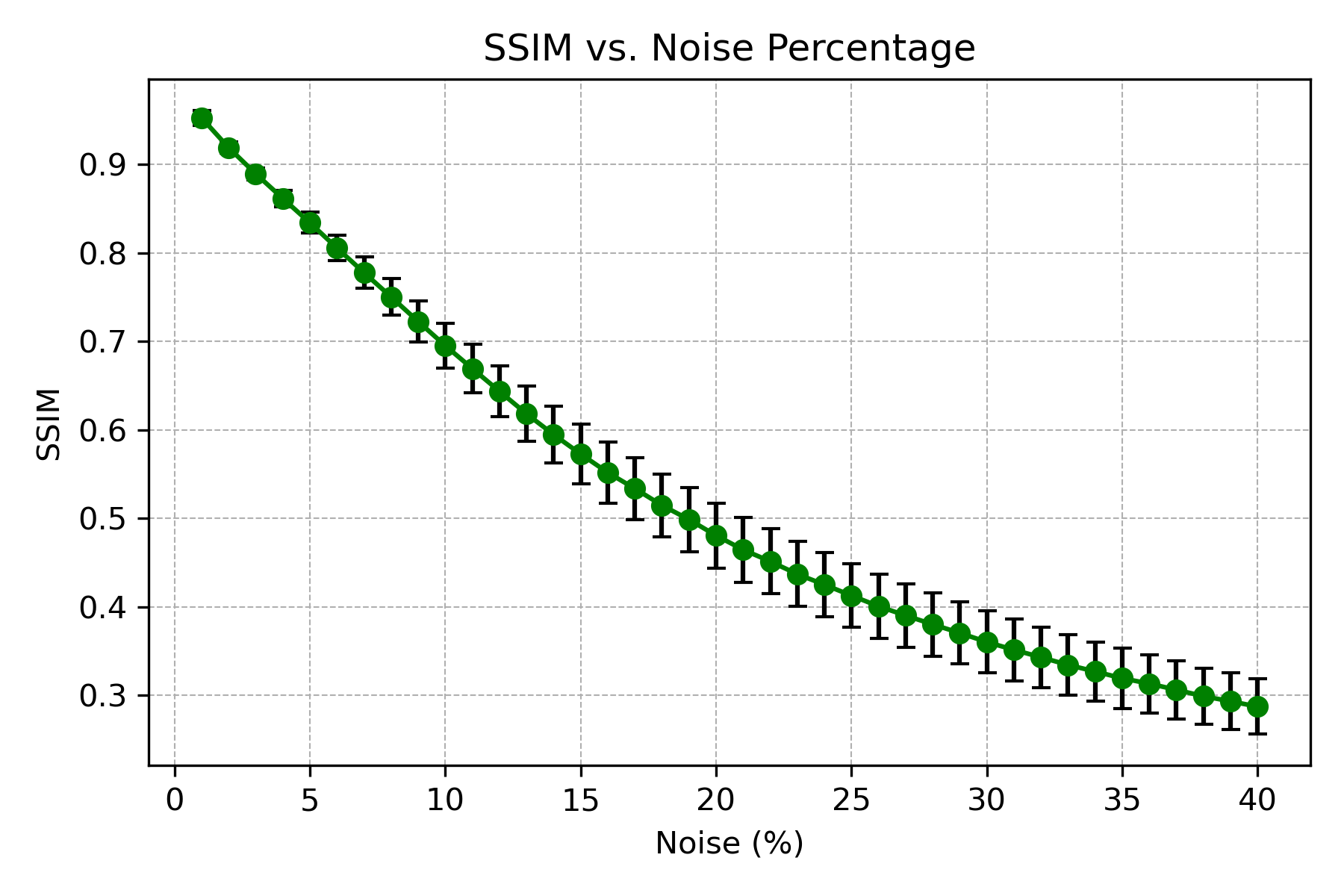}\label{fig:ssim}}
            \subfigure[]{\includegraphics[width=0.4\linewidth]{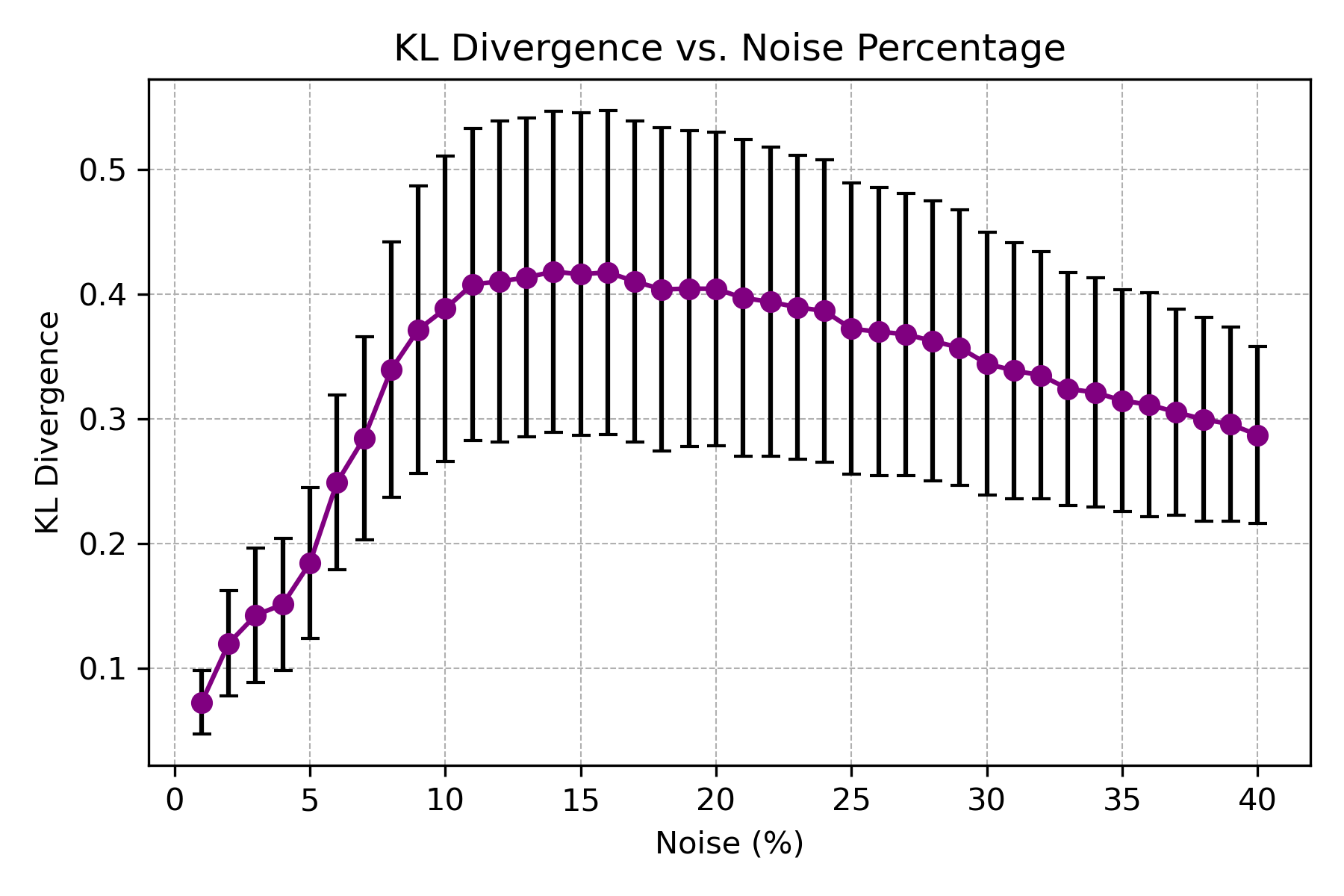}\label{fig:kldiv}}
            \caption{
                Variation of four evaluation metrics as a function of added Gaussian noise.
                (a) Mean absolute error.
                (b) Root mean squared error.
                (c) Structural similarity index.
                (d) Kullback–Leibler. Error bars indicate standard deviation for each test data.
            }
            \label{fig:noise_metrics}
        \end{minipage}
    }
\end{figure}

To quantify the impact of input noise on model performance, Gaussian noise was added to radiograph data in increments from 1\% to 40\% of the data range. For each noise level, WEPL maps were predicted for a test set of test data samples using the trained generator. The evaluation metrics MAE, RMSE, SSIM, and KL divergence were calculated at each noise level; results are summarized in Figure~\ref{fig:noise_metrics}, with error bars showing the standard deviation across all samples.

MAE increased from 0.025 at 1\% noise to 0.07 at 40\% noise, while RMSE rose from 0.028 to 0.09 over the same range. Both metrics followed a near-linear trend. SSIM decreased nonlinearly, from 0.95 at low noise to 0.3 at the highest noise level, indicating loss of structural similarity as noise increases. KL divergence rose rapidly from 0.06 at 1\% noise, peaking at approximately 0.42 around 10--15\% noise.

MAE and RMSE values remained below 5\% for noise levels up to approximately 10\%. However, as noise increased, SSIM decreased from 97\% to 70\%, while KL divergence increased from 6\% to 30\%. For higher noise, all four metrics reflected substantial degradation and increased variability, with KL divergence indicating that statistical agreement between prediction and reference distributions is no longer maintained beyond this point. 

\begin{figure}[!htbp]
    \centering
    \includegraphics[width=0.7\linewidth]{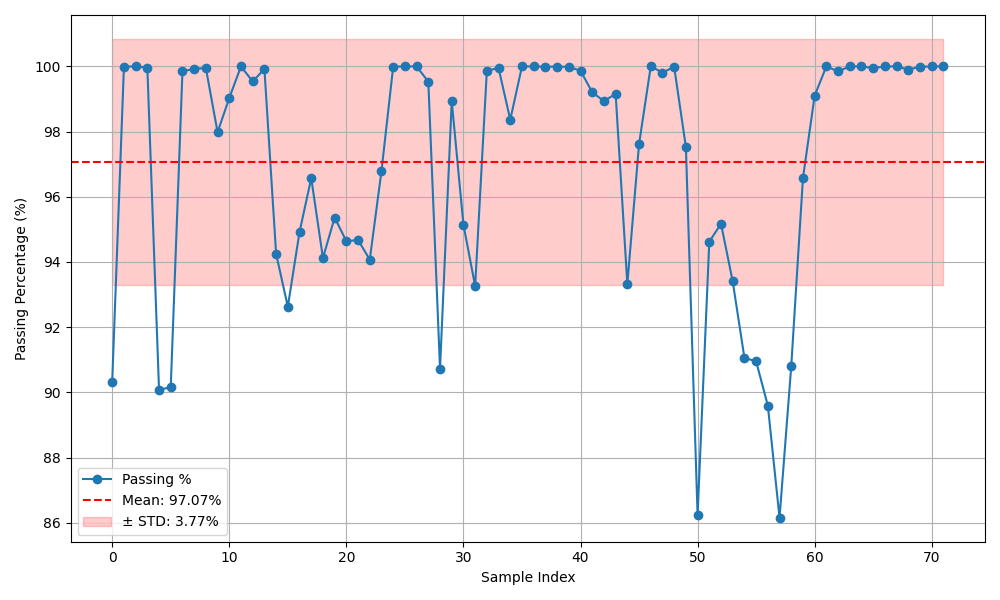}
    \caption{
        Gamma passing rate ($\gamma_{\mathrm{pRG}}$) for all test samples, with WEPL tolerance 2\% and distance-to-agreement tolerance 3~mm. The dashed red line shows the mean (97.07\%) and the shaded band indicates one standard deviation (3.77\%) about the mean.
    }
    \label{fig:gamma_passing_rate}
\end{figure}

Figure~\ref{fig:gamma_passing_rate} presents the gamma passing rate ($\gamma_{\mathrm{pRG}}$) for all test samples, calculated with a WEPL tolerance of 2\% and a spatial tolerance of 3~mm. Each point denotes the percentage of pixels in a given test sample for which $\gamma_{\mathrm{pRG}} < 1$, reflecting spatial and value agreement between the predicted and reference WEPL maps. The mean passing rate is 97.07\%, with a standard deviation of 3.77\%. Most samples exceed the 95\% passing threshold, indicating consistent model performance but samples with lower passing rates generally indicate either challenging anatomical regions or limitations in model generalization, which may guide future model improvements.

\section{Discussion}
This study introduces a direct deep learning framework for reconstructing WEPL maps from high-dimensional, energy-resolved proton radiographs, employing principal component analysis for dimensionality reduction and a conditional WGAN-GP model. The approach achieves strong performance on simulated head phantom data, as detailed in Table~\ref{tab:evaluation_metrics} and Figures~\ref{fig:comparison_wetSamples}--\ref{fig:gamma_passing_rate}, setting a promising benchmark for this application.

A key strength of the method is the use of PCA to compress 81-dimensional radiograph stacks into 16 principal components, retaining more than 99\% of relevant variance while suppressing noise and redundancy. This strategy allows efficient training of a 15-million-parameter model on standard GPU hardware, reducing both overfitting risk and computational burden, which is crucial for future clinical translation.

Combining adversarial, perceptual, and structural similarity losses enables recovery of both global and local anatomical features. The model demonstrates rapid convergence within 500 epochs and stable performance beyond 5,000 epochs, with no significant evidence of overfitting (see Figures~\ref{fig:epochvsmetrics} and~\ref{fig:normalized_loss}).
Compared to recent DRM-based and deconvolution approaches~\citep{Tsai2022DRM,Bentefour2016}, which report WEPL deviations in the 1--2\% range and gamma passing rates above 95\%, this model achieves similar accuracy, with greater flexibility and fewer handcrafted steps. Notably, PCA-based dimensionality reduction is applied here directly to WEPL mapping for the first time, leading to a more compact and robust model.

The DRM-based studies by~\citep{Tsai2022DRM,cho2025experimental} demonstrate superior numerical accuracy for WEPL prediction, with mean errors as low as 0.02~cm and gamma passing rates above 99\%. However, these results rely on structured phantoms and require multiple calibration and correction steps. Our deep learning approach, while yielding slightly higher errors using only linear principal components, offers a fully automated end-to-end solution directly applicable to complex, CT-based anatomical data, and provides greater flexibility for future clinical adaptation.
Despite strong performance, several limitations should be noted. The model was trained and tested only on simulated data from limited patient CTs, without accounting for detector noise, calibration errors, or inter-patient anatomical variation. While most test cases yield gamma passing rates above 95\%, a few challenging examples drop below 90\%. This highlights the need for larger, more diverse datasets and enhanced regularization or loss functions to improve robustness. The study currently addresses single-angle WEPL reconstruction only, with extension to full volumetric pCT and clinical dose calculation still required.

A significant practical limitation is the computational demand of data generation. Each sample involves 81 irradiations at different proton energies and $10^7$ protons per irradiation, across 72 projection angles. This makes simulation time-consuming and restricts dataset diversity. Expanding the dataset to multiple patients and anatomical sites, and using methods such as data augmentation, transfer learning, or synthetic data, will be necessary to support practical clinical deployment.
We acknowledge the sharp peaks observed in the SSIM and $\gamma_{pRG}$ passing rate distributions in Figures~\ref{fig:ssim1} and~\ref{fig:gI}. This effect arises from the high accuracy and consistency of our model on a homogeneous test set, where most samples consist of smooth, low-frequency anatomical structures. Because both SSIM and gamma passing rate are bounded metrics, with maximum values of 1 and 100\%, respectively, high-quality reconstructions naturally cluster near these upper limits. To avoid overinterpreting these results, we have reported standard deviations together with mean values. In future studies, we will include more structurally diverse and challenging samples to provide a stronger test of model robustness.

The model’s ability to capture statistical data features is also limited. While it effectively learns linear and nonlinear relationships for WEPL prediction, it is less robust under statistical variation and noise. For example, MAE and RMSE remain below 5\% for noise up to 10\%, but SSIM drops from 97\% to 70\%, and KL divergence rises from 6\% to 30\%. The sharp rise in KL divergence, which is particularly sensitive to distributional shifts, shows the challenge of modeling statistical variability. This suggests that future research should consider generative architectures such as variational autoencoders to improve robustness and fidelity under noisy conditions.

The results must be interpreted within the fundamental physical limits set by multiple Coulomb scattering~\citep{collins2020statistical}. No model, regardless of architecture, can recover anatomical detail that is physically lost due to scattering-induced blur. Our approach improves prediction fidelity and noise resilience but does not overcome this intrinsic spatial resolution barrier.
While PCA reduces complexity and noise, it is inherently linear and cannot capture nonlinear data relationships. Although principal components may visually represent complex patterns, fully leveraging anatomical structure may require nonlinear techniques such as autoencoders or attention-based models.
Interpretability remains a challenge. While principal components may encode physically meaningful features, their clinical interpretation requires further study. Applying explainable AI methods and uncertainty quantification will be important for clinical acceptance.
Selecting 16 principal components was based on preserving range-mixing features without excessive complexity. More systematic, quantitative optimization could refine this choice, possibly enabling lower-dose imaging protocols in the future.

\textbf{Future outlook.}
A key challenge for WEPL reconstruction is balancing the trade-offs between low and high proton beam energies: lower energies yield reduced noise but increased blurring, while higher energies improve spatial resolution at the cost of greater noise. These physical properties should inform the design of future deep learning models like physics informed deep learning models. Promising directions include leveraging the complementary strengths of low- and high-energy radiographs, potentially by extracting and fusing features from each regime through specialized or integrated networks. In addition, future research should validate the proposed approach on experimental and diverse clinical datasets, incorporate prior anatomical or multimodal imaging information, and explore advanced dimensionality reduction using nonlinear or learned embeddings such as variational autoencoders or attention-based mechanisms. Expanding to multi-angle volumetric WEPL mapping, integrating uncertainty quantification, and systematically investigating the relationships among imaging dose, number of projections, and reconstruction accuracy will be essential for optimizing performance and establishing clinical relevance

\section{Conclusion}

This study demonstrates a practical and accurate approach for direct WEPL reconstruction from high-dimensional proton radiographs using PCA-driven dimensionality reduction and a conditional GAN. The method achieves strong performance, with mean absolute errors near 0.02 and gamma index above 0.97, overcoming major limitations of traditional techniques. This framework sets the stage for high-fidelity WEPL prediction in proton therapy, with promising potential for future clinical application.

\section*{Declaration of competing interest}
The authors declare that they have no conflicts of interest or competing financial interests related to this work.

\section*{Acknowledgement}
This work was funded by the National Science and Technology Council, Taiwan,  NSTC~114-2121-M-182-001, 113-2121-M-182-001, 112-2314-B-182A-122-MY2, 114-2221-E-182-025, 113-2811-M-182-005 and by the Chang Gung Research Project CMRPG3K1911, CIRPD1I0023,  CMRPD2N0161 and BMRP736. We also thank the Radiation Research Core Laboratory,  Chang Gung Memorial Hospital, and Taiwan Space Agency for their assistance. 
\newcommand{\newblock}{}
\bibliographystyle{agsm} 
\small
\bibliography{verified1}
\normalsize

\end{document}